\documentclass[preprint,nofootinbib,superscriptaddress,preprintnumbers]{revtex4-1}

\usepackage{graphicx, hyperref, amsmath, amssymb, slashed, color, bbm}
\usepackage{hyperref}
\usepackage{breakurl}
\usepackage{array}
\usepackage{subfigure}
\makeatletter
\def\simgt{\mathrel{\lower2.5pt\vbox{\lineskip=0pt\baselineskip=0pt
           \hbox{$>$}\hbox{$\sim$}}}}
\def\simlt{\mathrel{\lower2.5pt\vbox{\lineskip=0pt\baselineskip=0pt
           \hbox{$<$}\hbox{$\sim$}}}}
\makeatother

\newcommand{\be}{\begin{equation}}
\newcommand{\ee}{\end{equation}}
\newcommand{\bea}{\begin{eqnarray}}
\newcommand{\eea}{\end{eqnarray}}

\begin{document}

\preprint{ MCTP-14-05}

\title{Monojet versus rest of the world I: $t$-channel Models}

\author{Michele Papucci}
\affiliation{Michigan Center for Theoretical Physics, University of Michigan, Ann Arbor, MI 48109, USA}
\author{Alessandro Vichi}
\affiliation{Berkeley Center for Theoretical Physics, 
  University of California, Berkeley, CA 94720, USA}
\affiliation{Theoretical Physics Group, 
  Lawrence Berkeley National Laboratory, Berkeley, CA 94720, USA}
\author{Kathryn M. Zurek}
\affiliation{Michigan Center for Theoretical Physics, University of Michigan, Ann Arbor, MI 48109, USA}

\begin{abstract}

Monojet searches using Effective Field Theory (EFT) operators are usually interpreted as a robust and model independent constraint on direct detection (DD) scattering cross-sections.  At the same time, a mediator particle must be present to produce the dark matter (DM) at the LHC.  This mediator particle may be produced on shell, so that direct searches for the mediating particle can constrain the effective operator being applied to monojet constraints.  In this first paper, we do a case study on $t$-channel models in monojet searches, where the (Standard Model singlet) DM is pair produced via a $t$-channel mediating particle, whose supersymmetric analogue is the squark.  We compare monojet constraints to direct constraints on single or pair production of the mediator from multi-jets plus missing energy searches and we identify the regions where the latter dominate over the former.   We show that computing bounds using supersymmetric simplified models and in the narrow width approximation, as done in previous work in the literature, misses important quantitative effects. We perform a full event simulation and statistical analysis, and we compute the effects of both on- and off-shell production of the mediating particle, showing that for both the monojet and multi-jets plus missing energy searches, previously derived bounds provided more conservative bounds than what can be extracted by including all relevant processes in the simulation.  Monojets and searches for supersymmetry (SUSY) provide comparable bounds on a wide range of the parameter space, with SUSY searches usually providing stronger bounds, except in the regions where the DM particle and the mediator are very mass degenerate. The EFT approximation rarely is able to reproduce the actual limits. In a second paper to follow, we consider the case of $s$-channel mediators.

\end{abstract}

\maketitle

\tableofcontents

\section{\textbf{Introduction}}

Monojet plus missing transverse energy (MET) searches at colliders~\cite{Nachtmann:1984xu,Dicus:1989gg,Brignole:1998me} have become a powerful tool for constraining dark matter (DM) scattering cross-sections off of nucleons in direct detection (DD) experiments.  The basic idea is to make use of crossing symmetry to relate the scattering cross-section of DM off of nucleons to the production of DM at colliders.  The DM is produced with initial state radiation (ISR), whether that be a $Z$~\cite{Petriello:2008pu,Carpenter:2012rg}, photon~\cite{Birkedal:2004xn,Gershtein:2008bf,Nelson:2013pqa}, jet~\cite{Goodman:2010yf,Goodman:2010ku,Bai:2010hh,Fox:2011fx,Fox:2011pm,Fox:2012ee,An:2012va}, top~\cite{Lin:2013sca} or Higgs~\cite{Davoudiasl:2004aj,Carpenter:2013xra}, allowing a measurement of the missing transverse energy (MET) carried by the DM.  The strongest constraints are derived from a jet in the initial state.  Within the context of an Effective Field Theory (EFT), the DM production cross-section has a unique mapping onto a DD cross-section for a given operator, which has been used to place constraints on DD of DM~\cite{Goodman:2010yf,Goodman:2010ku,Bai:2010hh,Fox:2011fx,Fox:2011pm,Fox:2012ee}.  

New dynamics must be involved in generating the effective operator, with the simplest case being a single particle mediating the interaction at tree level. That particle will be coupled to quarks and/or gluons, and, if kinematically possible, may be resonantly produced at the collider. In this regime, the EFT is no longer valid~\cite{Dreiner:2013vla,Busoni:2013lha,Buchmueller:2013dya,Chang:2013oia,An:2013xka,Bai:2013iqa,DiFranzo:2013vra} and the question is whether the actual limits can be stronger or weaker. The regions with the largest cross-section are both at low center of mass energy (due to parton distribution function effects), and on resonance~\cite{Petriello:2008pu,Gershtein:2008bf,Dreiner:2013vla,Busoni:2013lha,Buchmueller:2013dya,Chang:2013oia,An:2013xka,Bai:2013iqa,DiFranzo:2013vra}. On the other hand, production at low $\sqrt s$ is the most contaminated by SM backgrounds. 

Crossing symmetry connects DM production at a hadron collider and DD cross-sections whenever both the partons and the DM involved in the scattering are close to being on-shell, with very little sensitivity to the details of the interaction. On the other hand, requiring an extra hard jet from initial state radiation, puts one of the partons producing the DM pair highly off-shell. Therefore, as in Deep Inelastic Scattering, the production cross-section becomes sensitive to any mass scale of the order of the momentum transfer. Since the monojet analyses at the LHC happen to provide bounds on EFT scales which are of the same order as the MET cut in the search, as we will see in Sec.~\ref{sec:model and searches}, the searches will always be sensitive to the internal structure of the effective operator unless a separation between the EFT scale and the mass of the particle mediating the interaction is achieved.  Since $\Lambda = m_M/g_M$, where $m_M$ and $g_M$ are the mass and coupling of the mediator to the DM, this separation of scales requires the introduction of a large $g_M$.
 
In fact, by comparing the EFT to a UV-complete theory including both on- and off-shell production of the mediating particle, one can easily see the constraints from the UV-complete theory may be stronger or weaker than those extracted from the EFT.  If the mediator is accessible and sufficiently heavy (with a mass of the order of the MET cut applied in the monojet search) the actual bounds will be stronger than the EFT limit on account of resonant production~\cite{Fox:2011pm}. In this sense, the collider constraints are conservative. On the other hand if the mediator (and the DM) is light the signal will appear mostly in the region contaminated by the SM and the constraints from an EFT analysis are overly strong~\cite{An:2012va}. 

Perhaps, more importantly, however,  the presence of a dynamical mediator particle opens the possibility of directly searching for it in other final states. Such searches may have stronger impacts on the DD limits than the monojet searches.  Since the mediator couples to quarks in the initial state and/or is a colored particle, this leaves open the possibility that the mediator, rather than going only to DM, may decay to dijets (as in the $s$-channel case) or may be pair produced and be detected in multi-jet plus MET final states (for a $t$-channel mediator).   

In this paper we take a global view of the constraints in various limits of the mediator mass and its width, taking as a case study $t$-channel DM production.  In the case of $t$-channel models, the canonical example is DM pair production through a $t$-channel squark.  The constraints from monojet searches can be compared against squark pair production, with each squark decaying to the DM plus a jet, so that these models can also be constrained by looking at final states containing at least two jets plus MET. The effect of multi-jet plus MET searches for a squark mediator was assessed in~\cite{Chang:2013oia,An:2013xka,Bai:2013iqa,DiFranzo:2013vra}.  Quantitatively, our results differ substantially from the results of these previous works.  Previous works simulated squark pair production and extracted the constraint on the coupling from the relevant squark-neutralino simplified model interpretations of the jets+MET searches~\cite{TheATLAScollaboration:2013fha,ATLAS:2012ona,ATLAS-CONF-2012-033,TheATLAScollaboration:2013aia,CMS-PAS-SUS-13-012,CMS:zxa,CMS:2012kba,Chatrchyan:2012lia}. This is expected to be valid for DM couplings parametrically smaller than the strong coupling constant and in the narrow-width approximation (NWA). By contrast, we find that in most of the parameter space the limits obtained constrain ${\cal O}(1)$ or larger values of DM coupling. Therefore it is necessary to recast the jets+MET searches with a proper simulation to capture all the relevant effects necessary to extract the correct constraints on the cross-section.  We find that, in many regions of the mediator mass and DM mass parameter space, the jets+MET constraints close the parameter space where monojets by themselves are unconstraining. 

In a companion paper, we consider $s$-channel DM production in monojet searches, comparing the monojet to dijet constraints on the mediator particle.   A canonical example is the $Z'$ mediator, where monojet constraints on DM pair production can be compared against dijets when the $Z'$ decays back to SM quarks. The impact of dijet searches was assessed already in~\cite{An:2012va} for the specific cases of the $Z'$ mediator and in~\cite{Dreiner:2013vla} for the case of an effective operator.  In our analysis we will again simulate both on and off-shell production of the mediator particle to both DM and to dijets and compare against existing constraints from both dijet and jets+MET. 

We focus on the $s$- and $t$-channel cases because they represent tree level mediation of singlet DM. Such cases generically offer the best constraints from direct DM (monojet) searches compared to mediator searches. In fact in all the other cases other relatively light non-singlet particles are present (either its partners if the DM is charged under the weak interactions or the mediators running in loops, whose masses will be parametrically lower than the effective mediator scale probed by direct searches).   This in turn will generically provide better search handles than the monojets searches. There are of course counter-examples to this argument, the most important being the case of light electroweak DM in the degenerate region, where adding additional electrically charged states (charginos) quasi-degenerate with the DM slightly boosts the monojet limits without providing extra handles for different collider searches.  \\
 \indent We map our results to the direct detection (DD) $m_{DM}-\sigma_{DM}$ plane.  Because the constraints depend on the mediator mass regime, we obtain bands (as opposed to lines) of constraints which can be compared against the results of DD experiments.  
The EFT constraints usually fall in the middle of the band: in the case of resonant production, the EFT constraints are weaker than the true constraints and represent a conservative bound;
on the other hand, when the mediator is fairly light or broad, the EFT constraints are stronger than the true constraints.  As a result, \emph{while monojet constraints on DD are relatively model independent in the EFT regime} (which is not entered until the mediator is above 3 TeV in the $s$-channel case, and 1 TeV in the $t$-channel case and the DM is parametrically lighter), \emph{they rarely represent the true constraints.}   \\
\indent
The outline of this paper is as follows.  In the next section we review the model we are constraining, as well as monojet searches for DM and the complementary multi-jet plus MET searches at the LHC. We briefly review the relevant scaling of cross-section necessary to interpret the results, and describe our set-up for analyzing the various LHC searches. Further technical details are given in Appendix~\ref{appendix:CLS}, along with a summary of the ATLAS and CMS monojet and jets+MET searches that we utilize. In Section~\ref{sec:monojet versus dijet} we present our results, comparing jets+MET searches against the monojet searches at the LHC.  We map our results to the DD plane, and we consider the compressed spectrum case.  Our conclusions are in Section~\ref{Sec:Conclusions}.

\section{\textbf{The Model and the Searches}}
\label{sec:model and searches}

The general class of models we are interested in is characterized by the presence of a DM candidate, which we will denote $\chi$, with mass $m_{DM}$, whose interactions with SM particles  are mediated by the exchange of a heavier state in the $t$-channel. A prototypical example of such model is for instance the MSSM where only the squarks and the neutralino are light, with the latter being the lightest one. 

More concretely we will study a model defined by the Lagrangian
\be
\mathcal L = \mathcal{L}_{SM} + g_{M}\sum_{i=1,2} \left(\widetilde{Q}_L^i \bar{Q}^i_L +  \tilde{u}^i_R \bar{u}^i_R + \tilde{d}^i_R \bar{d}^i_R \right)\chi + \text{mass terms} +c.c. 
\label{eq:lagrangian}
\ee 
where $Q_{L}^i, u_R^i, d_R^i$ are the usual SM quarks, $\widetilde{Q}_L^i, \widetilde{u}_R^i, \widetilde{d}_R^i $ correspond to the respective ``squarks''\footnote{We will use the terms squark and mediator interchangeably.}, and $i$ represents an index running on the fist two flavor families, since we will look at signals not involving the third generation. The fermion $\chi$, contrary to the SUSY case, is taken to be Dirac.  In order to maintain maximal flavor symmetry we took the squark masses and the couplings with DM to be equal. In general, as is well known, consistent with flavor, $\tilde Q_{L}^{a}$, $\tilde u_{R}^{a}$, $\tilde d_{R}^{a}$ or any combination of them may be present with the matrices $M_{Q,U,D}$ being different as long as they are proportional to the identity in flavor space. LHC limits are sensitive to it~\cite{Mahbubani:2012qq}. For simplicity we will take the squark masses to be degenerate and focus on two different extreme cases: 1) all squarks are present or 2)  only $\tilde d_{R}^{a}$ are present. They respectively maximize and minimize the squark production cross-section by multiplicity (and to a lesser extent, given the absence of a ``gluino'', parton distribution function effects). 

The above Lagrangian induces a minimal decay width for each squark given by the expression
\be
\Gamma_M^{min} = \frac{g_M^2 m_M}{16 \pi} \left(1-\frac{m_{DM}^2}{m_M^2}\right)^2,
\label{Gammamin}
\ee
where $m_M$ is the mediator mass. Clearly the squark width can be taken to be larger allowing the presence of additional states to which they can decay. These additional states may be constrained by other LHC searches than those considered here. We will be agnostic about the contribution to the mediator decay width from additional states given its model dependence.

\begin{figure}[htb]
\centering

\subfigure[]{\includegraphics[scale=0.30]{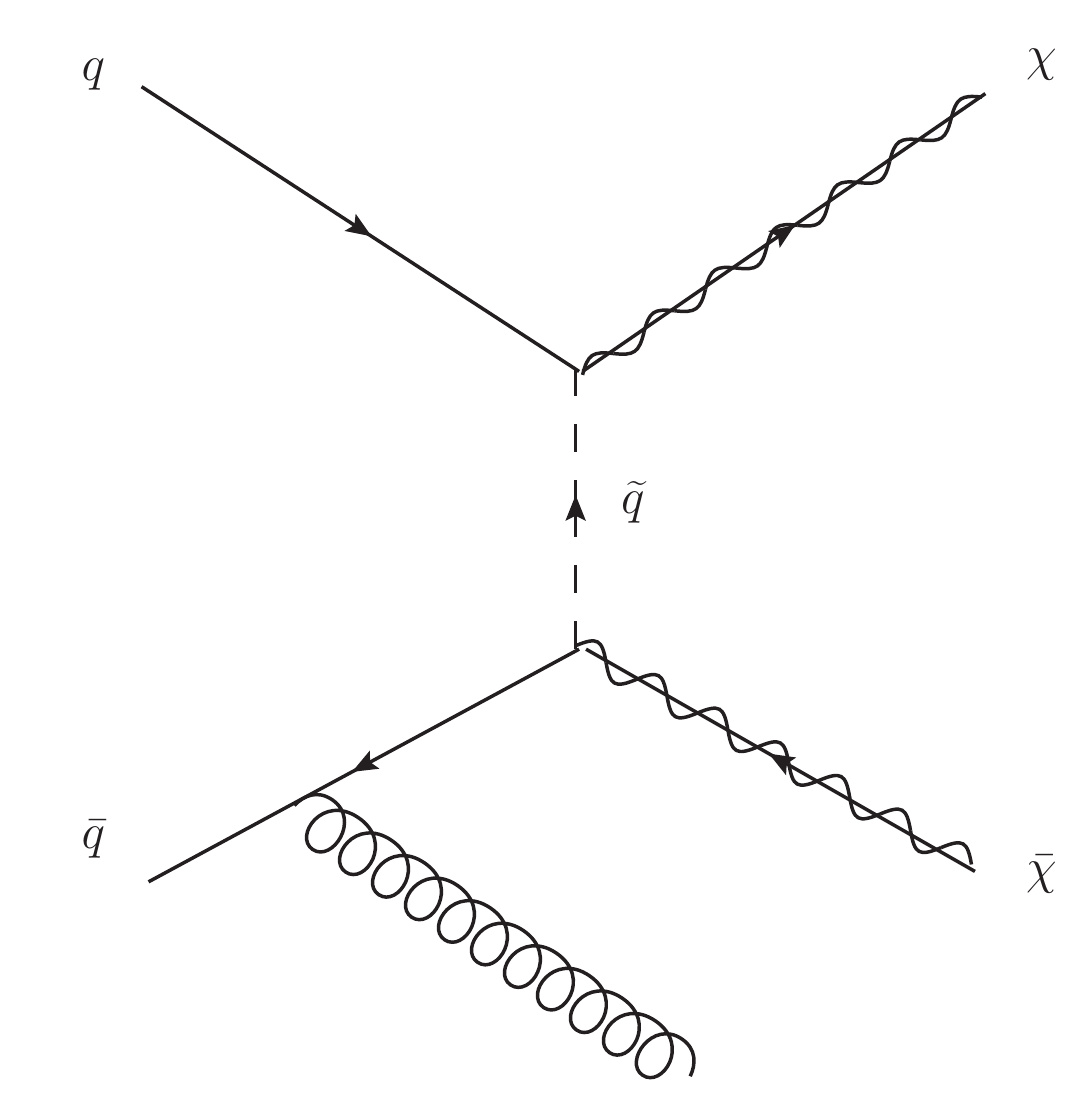}}
\subfigure[]{\includegraphics[scale=0.30]{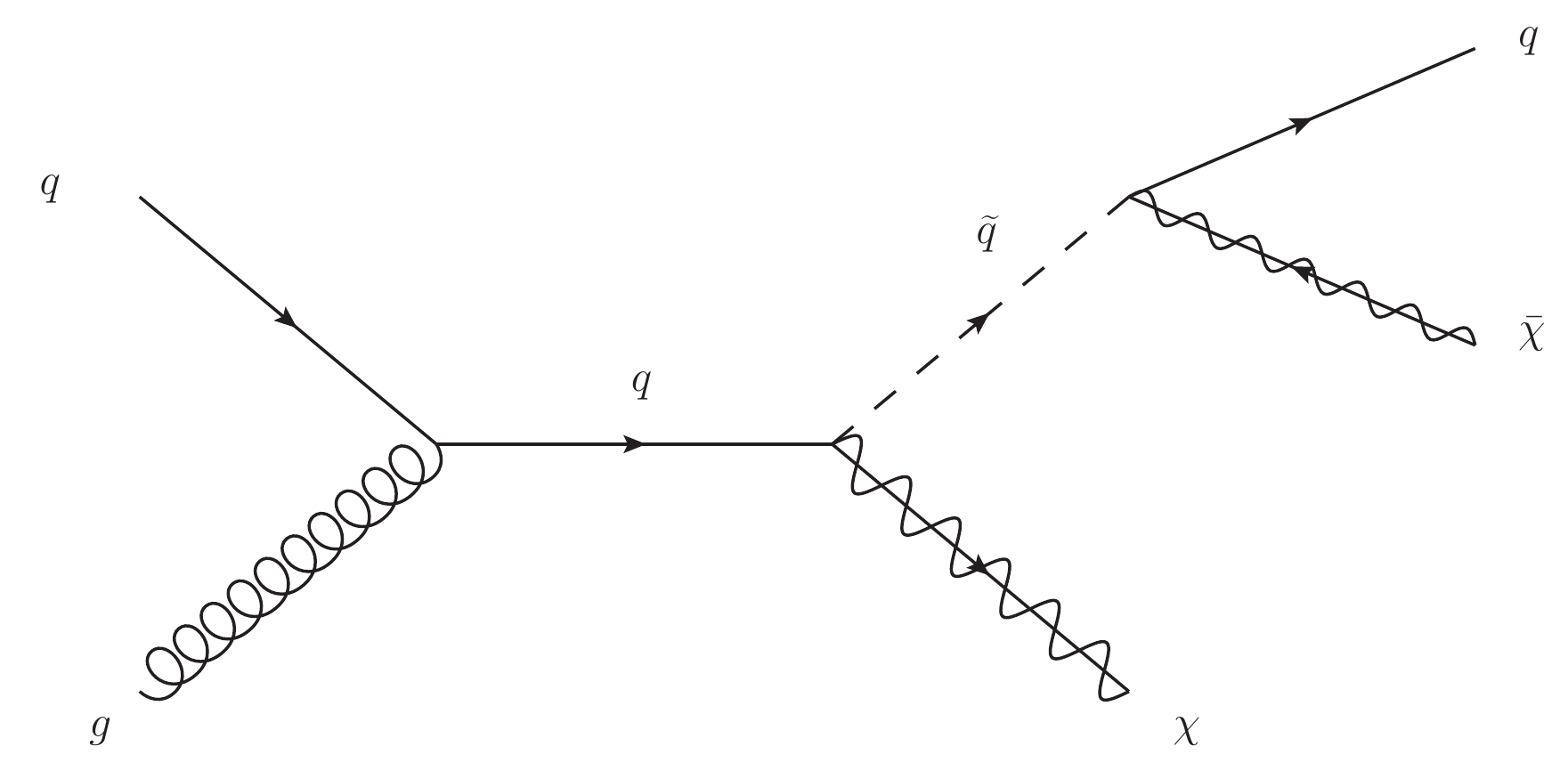}}
\subfigure[]{\includegraphics[scale=0.30]{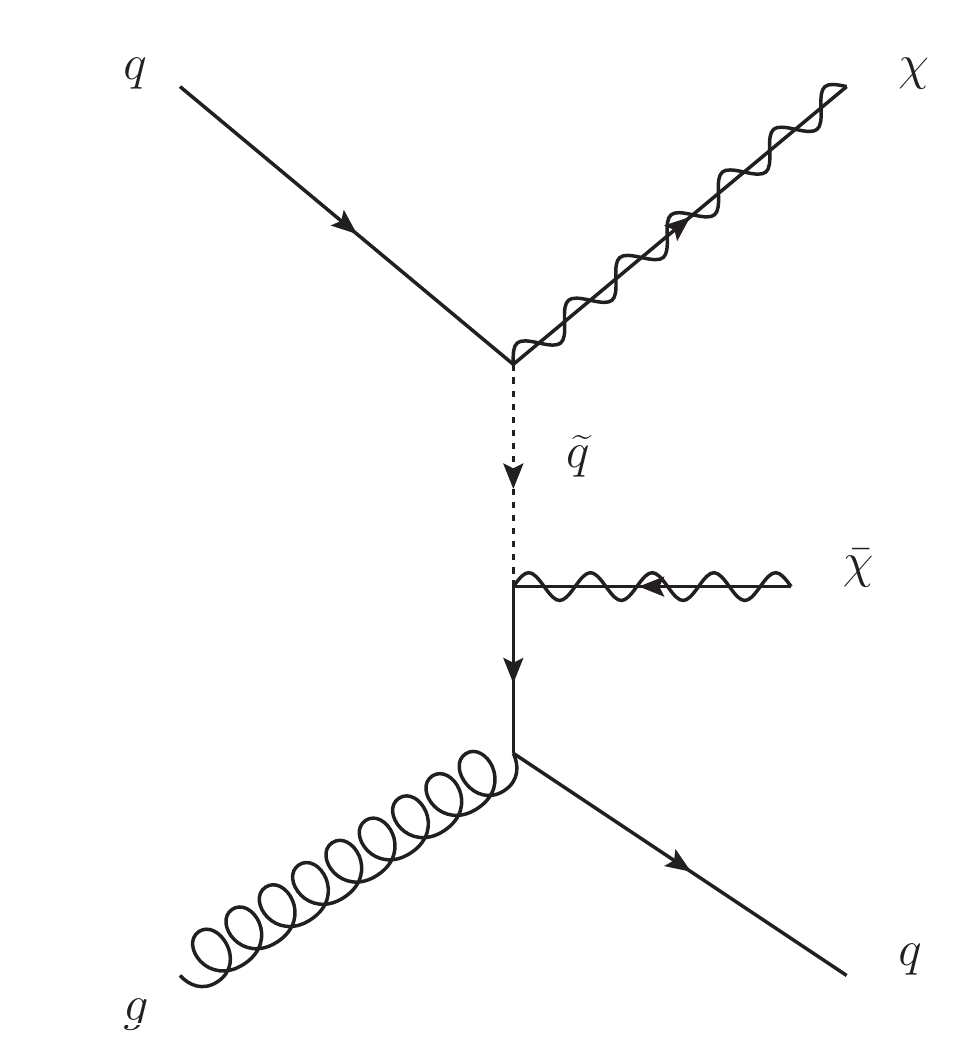}}\\
\subfigure[]{\includegraphics[scale=0.30]{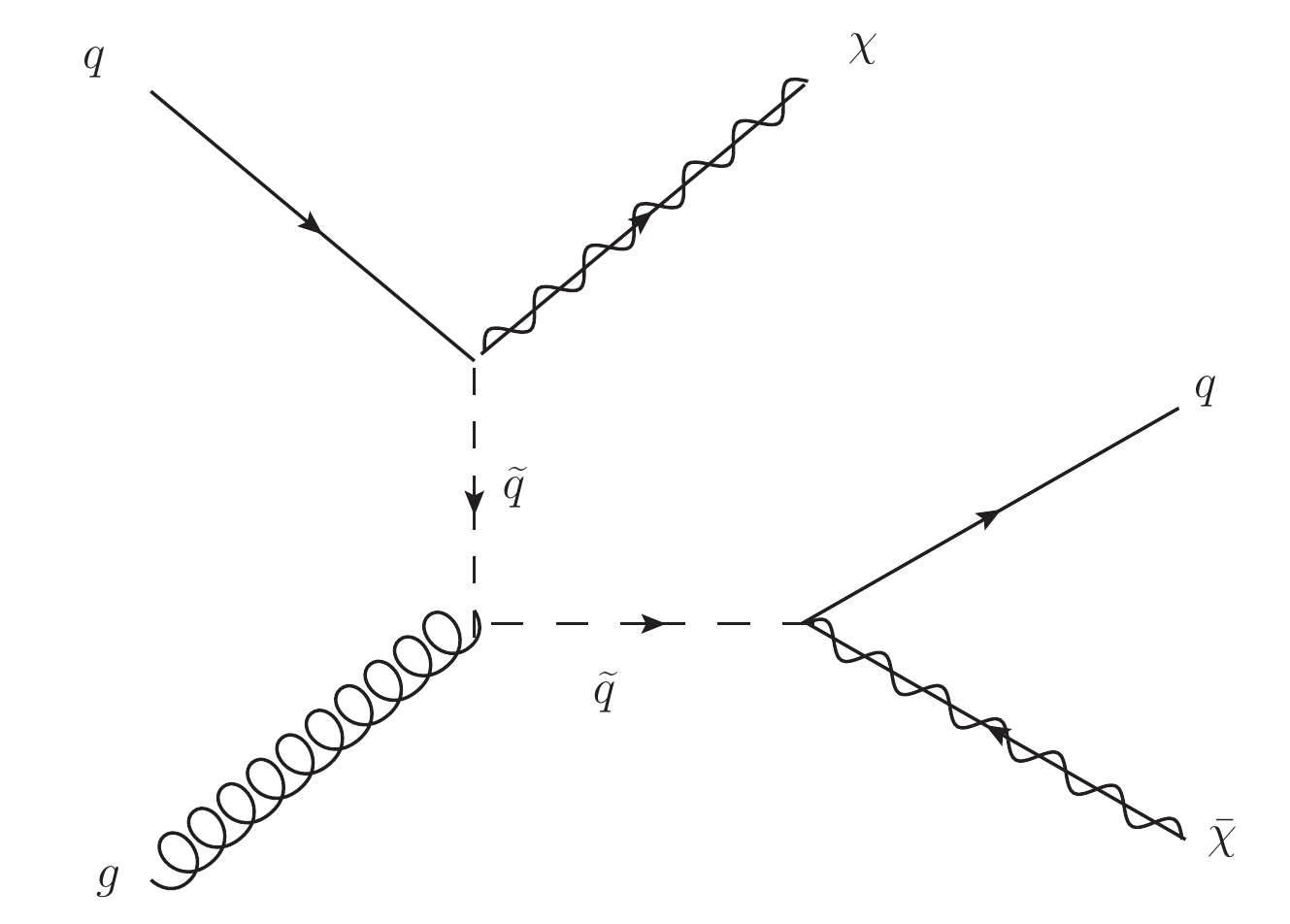}}
\subfigure[]{\includegraphics[scale=0.30]{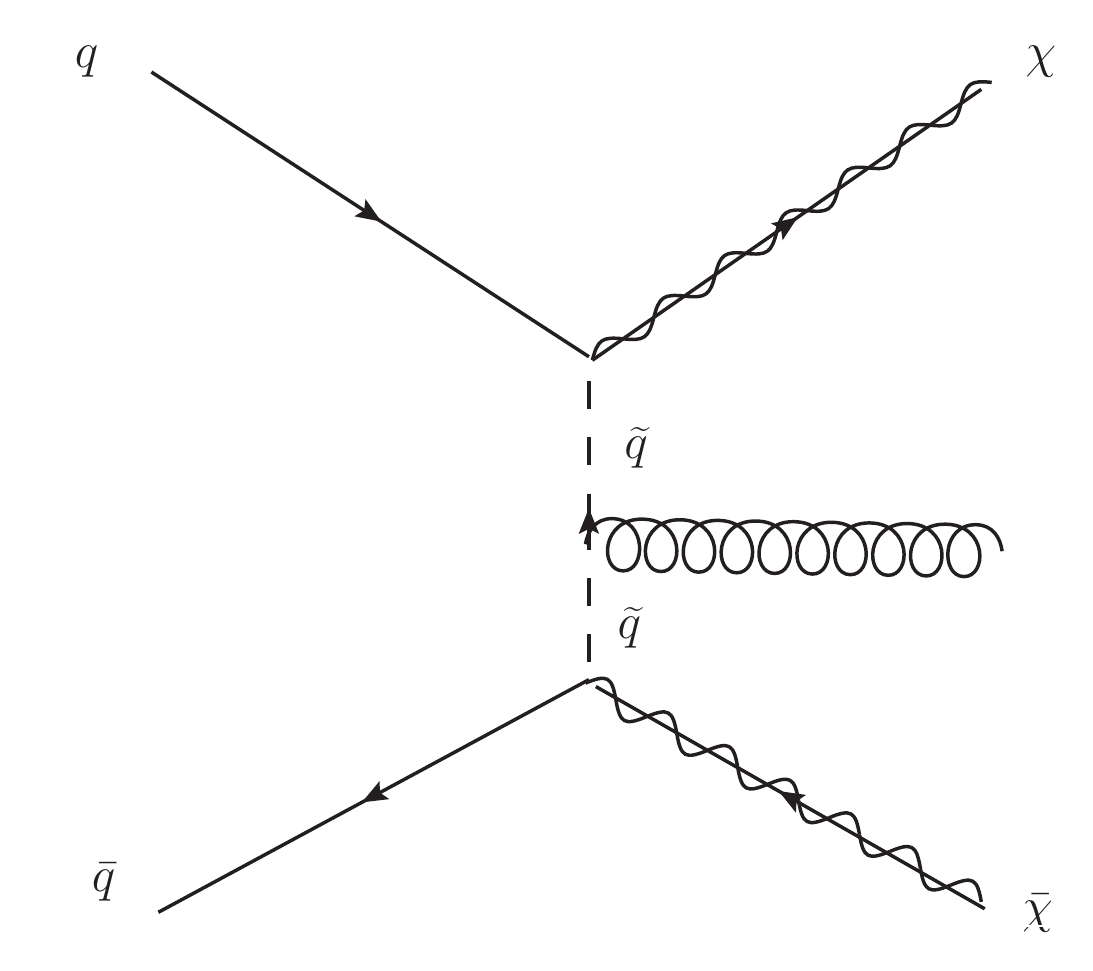}}
\caption{ Sample Feynman diagrams for monojet $t$-channel.  In the EFT limit only the first row dominates.} 
\label{fig:feynmanDM}
\end{figure}

The standard procedure adopted until recently to constrain the model at a collider was to extract limits on the effective operator mediating the DM-SM interaction from its predominant signature, which is a monojet and missing energy.   The constraints from monojet searches can be simply applied to DM DD in the limit that the mediator mass, $m_M$, is well above the typical production energies at the collider, $m_M \gg \hat{s}$.  The typical diagrams for DM pair production in association with a single jet are shown in Fig.~\ref{fig:feynmanDM}. By taking the heavy mass mediator limit, only diagrams (a-c) contribute and are encoded in a dimension six operator with a gluon attached to one of the external legs, while (d-e) contribute at dimension eight.  In this case, the collider DM production cross-section scales roughly as
\begin{eqnarray}
\sigma_t & \sim  \frac{g_M^4}{m_M^4} \equiv \frac{1}{\Lambda_{DD}^4}.
\label{EFTcross-section}
\end{eqnarray}
In this limit, $\Lambda_{DD}$ maps uniquely to a constraint on the direct detection cross-section, $\sigma_{DD}$, which scales precisely the same way, so that monojet constraints can be compared uniquely to the results from direct detection experiments. However, as already explained in the introduction, when the momentum transfer ({\em i.e.} the off-shellness of one of the quarks interacting with the DM) in diagrams (a-c) becomes of the order of the squark mass, the cross-section will be dependent on the full squark propagator structure. Since the momentum transfer is controlled by the largest between the $p_{T}$ cut on the mono-jet and the MET cut, for the EFT to be valid $m_{M}\gg {\rm max}\left(p_{T}^{j},\slash E_{T}\right)$. On the other hand current LHC searches happen to be sensitive to values of $\Lambda_{DD}$ not too far from the MET cut, so that the EFT limit requires both $g_{M}$ and $m_{M}$ to be large.

\begin{figure}[htb]
\centering
\subfigure[]{\includegraphics[scale=0.30]{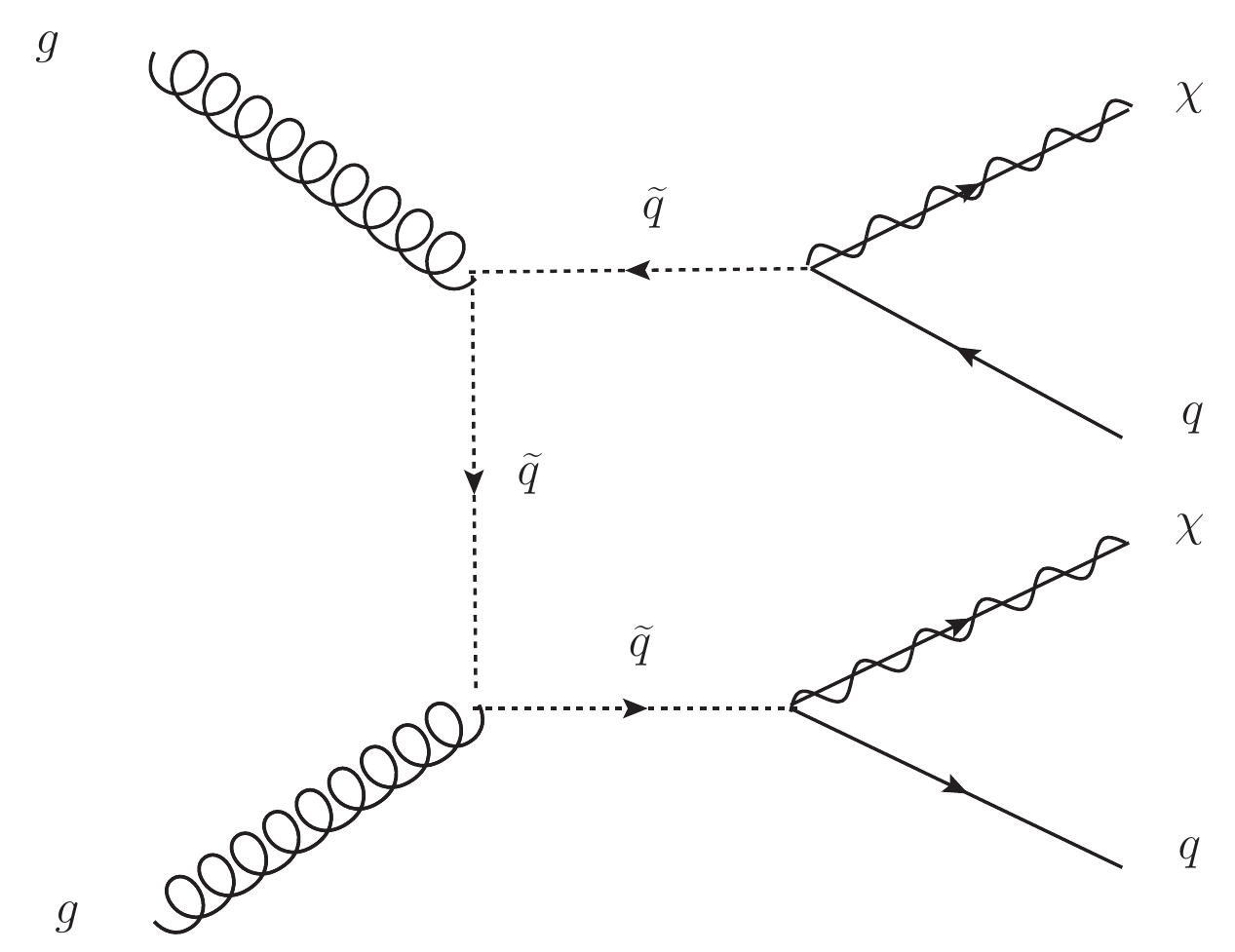}}\quad
\subfigure[]{\includegraphics[scale=0.30]{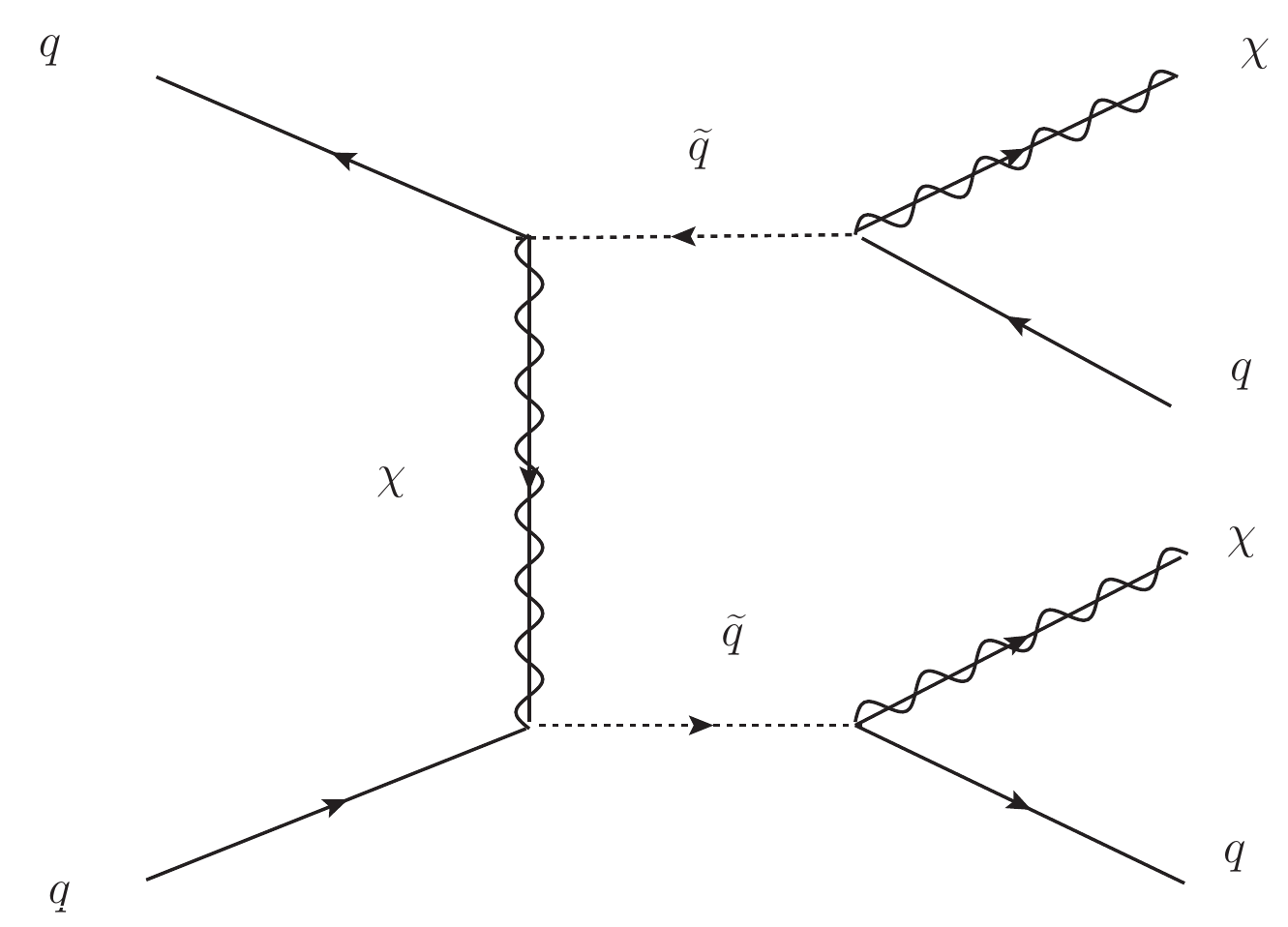}}\quad
\subfigure[]{\includegraphics[scale=0.30]{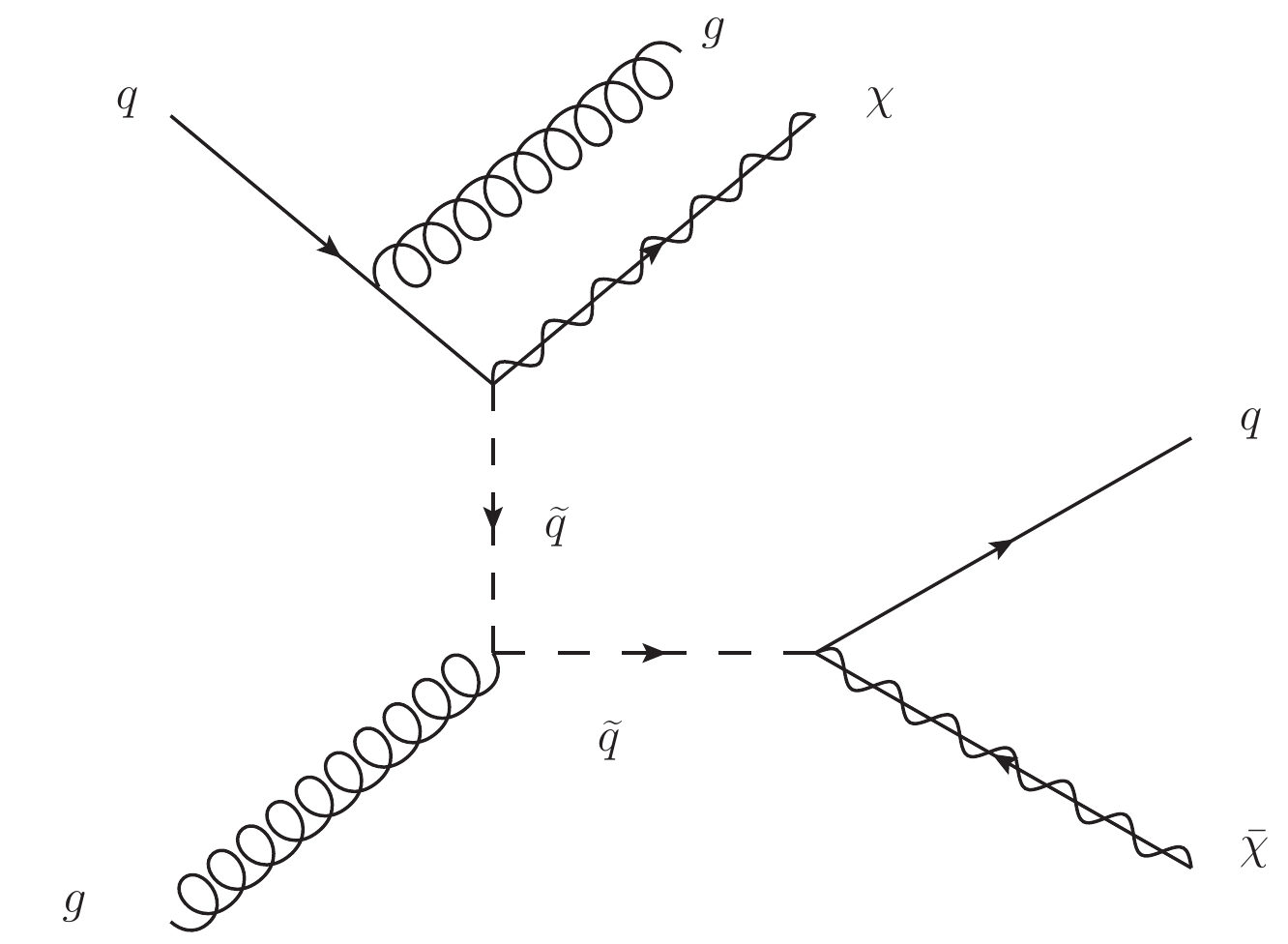}}\\
\subfigure[]{\includegraphics[scale=0.30]{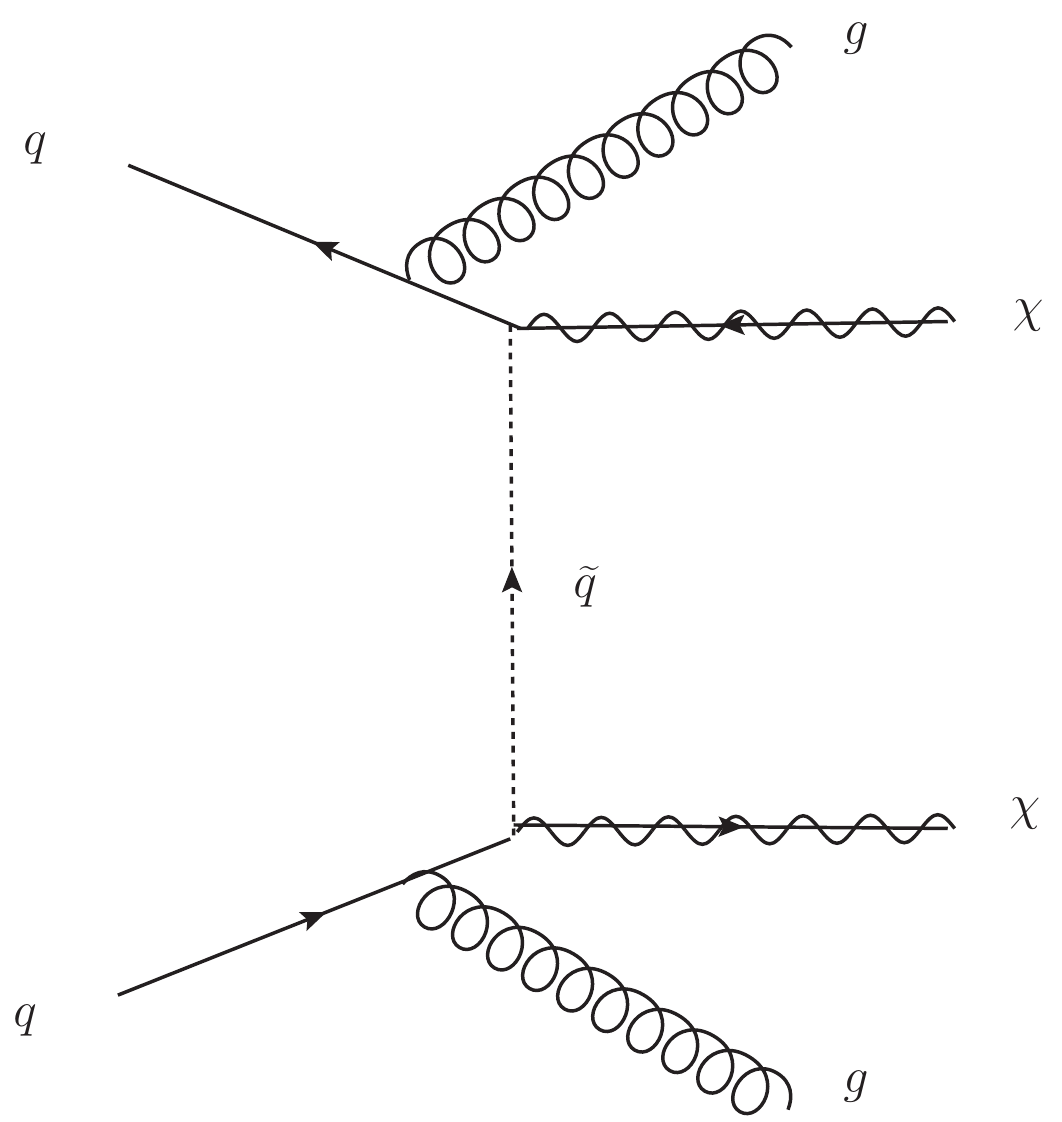}}\quad
\subfigure[]{\includegraphics[scale=0.30]{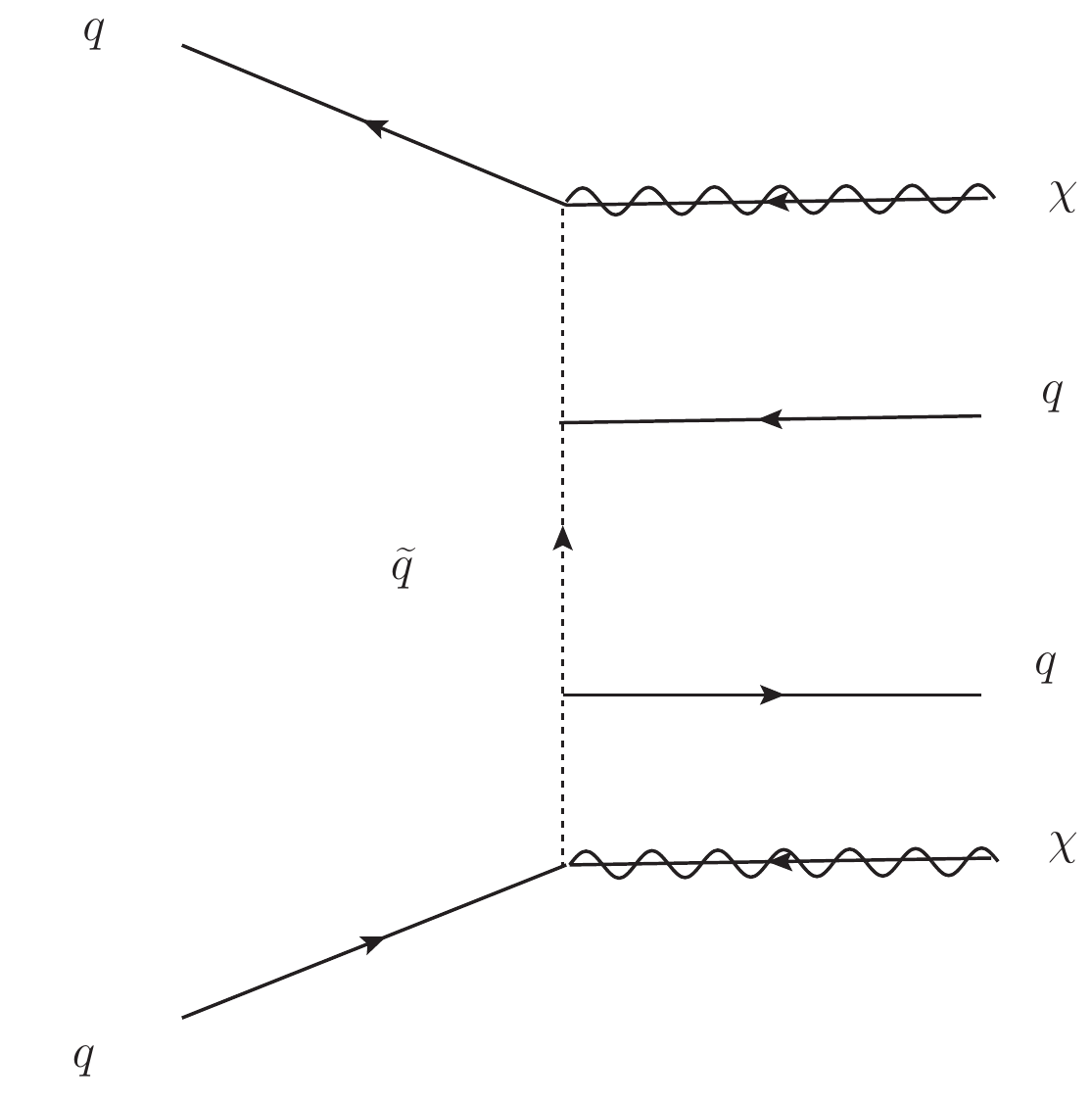}}
\caption{ Sample Feynman diagrams for dijets and MET.} 
\label{fig:feynman_2j}
\end{figure}

In the model considered the mediator couples directly to quarks, which means that it may decay back to quarks in association with a DM particle. This can already be seen in diagram (b)  of Fig.~\ref{fig:feynmanDM}, which can be interpreted as squark-DM associated production. The main point is to quantitatively compare the relative strength of monojet searches to a direct search for the mediator particle. Hence we consider a second final state, namely two jets and missing energy. In Fig.~\ref{fig:feynman_2j} we collected some of the most relevant contributions at parton level.  This dijet final state, and the comparison to the monojet final state, has already been considered in the literature~\cite{An:2013xka,Chang:2013oia,Bai:2013iqa}.  There is, however, an important quantitative difference between our treatment and previous treatments: previous works simulated only on-shell squark production, extracting the constraint on the size of the coupling by computing the on-shell squark pair production rate using MadGraph, and then comparing it to the quoted constraint on the SUSY squark-neutralino simplified model results presented in the experimental analysis.  By contrast, we perform a full simulation for a multi-jet plus MET final state including both on and off shell squarks, including interference with the Standard Model.  In the next subsections, we will see the reason that the different approaches yield different results for both monojet and jets+MET searches, and in which regions of parameter space these differences are most important.

\subsection{Monojet: importance of additional jets}\label{sec:mono}

Before presenting our bounds we describe the procedure we adopted and its differences with the common methodology adopted in the literature. 


In order to utilize monojet searches we simulated events of DM production in association with jets using MadGraph5~\cite{Alwall:2011uj}. Samples with different jet multiplicities were merged in the MLM scheme~\cite{Mangano:2006rw}. Up to two extra partons were included in the hard process. The events were then showered using Pythia6~\cite{Sjostrand:2006za}, and then passed through Atom\cite{Atom}, a software package for reinterpreting BSM searches based on Rivet~\cite{Buckley:2010ar}, and finally Fastjet~\cite{Cacciari:2011ma} for jet clustering. Both CMS~\cite{CMS-PAS-SUS-13-012,CMS:zxa,CMS:2012kba,Chatrchyan:2012lia,CMS-PAS-EXO-12-048,Chatrchyan:2012me} and ATLAS~\cite{TheATLAScollaboration:2013fha,ATLAS:2012ona,ATLAS-CONF-2012-033,TheATLAScollaboration:2013aia,ATLAS:2012zim,ATLAS:2012ky} mono jet analyses are cut-and-count based and consist on signal regions differing by $p_T$ and missing $E_T$ cuts. We validated Atom results against the various results provided in the experimental paper. Limits are set independently for each signal region and the upper bound on the number of signal events is provided by the collaborations, so that no further statistical analysis is required. For each analysis we choose the signal region providing the best expected limit on our model. Varying the value of $g_M$ in the simulation we found the maximal allowed couplings compatible with observations.
One should note that as the coupling increases, the width of the squarks must taken at least as large as $\Gamma_M^\text{min}$, according to Eq.~\ref{Gammamin}.  

\begin{figure*}[!h]
\begin{center}
\subfigure[]{\includegraphics[scale=0.7]{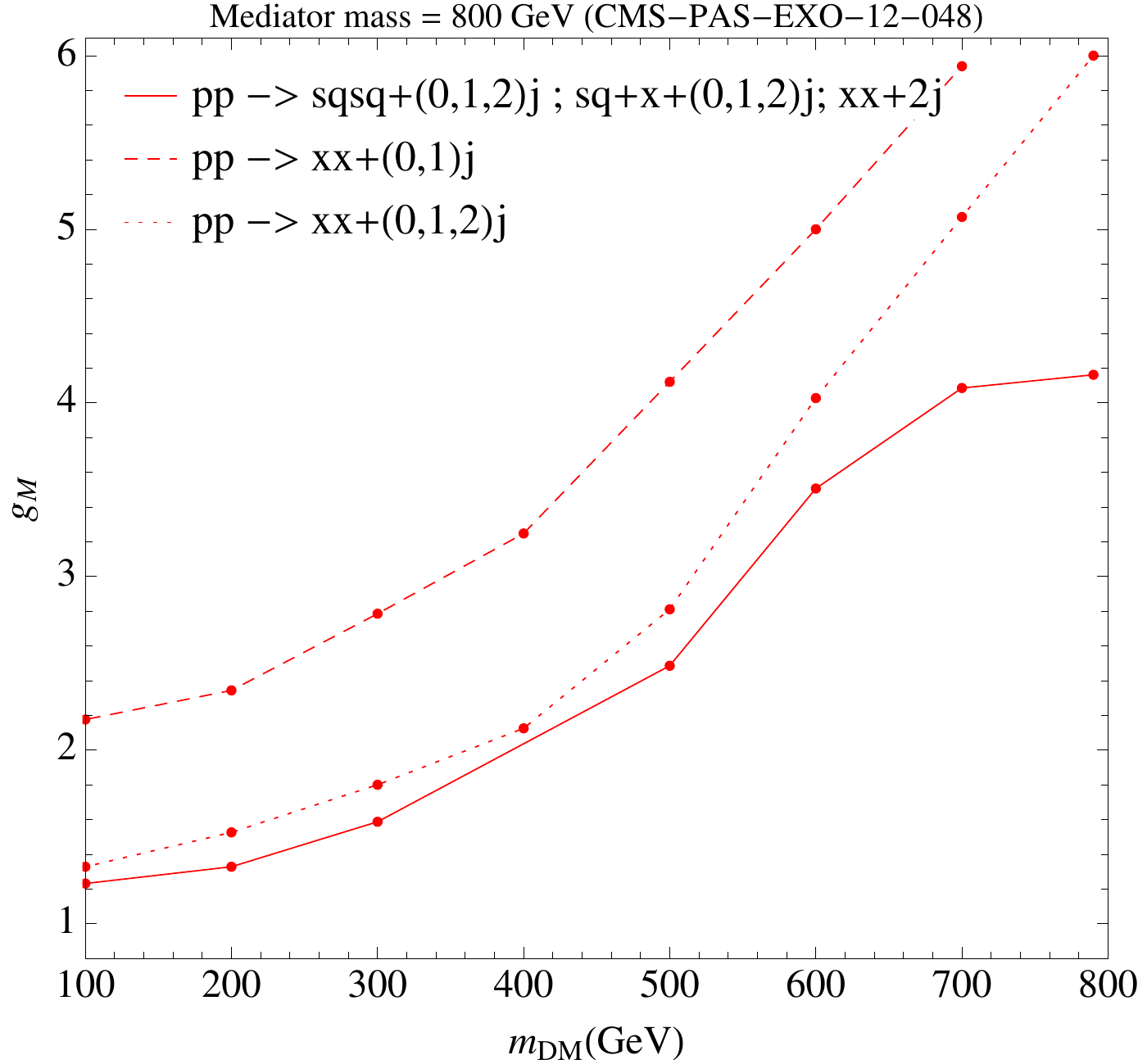}}
\caption{Exclusion limit from monojet searches obtained simulating events according to the three methods described in the text. The dashed line corresponds to MET plus one jet; the dotted line contains events with MET and up to 2 jets, showing the importance of additional jets even for monojet searches;  the continuous line is the combination of three split samples as described in the text.  In the compressed region, $m_M - m_{DM} \gtrsim 300 $ GeV, the additional jets obtained in the split sample are crucial for obtaining the correct constraint on $g_M$. }
\label{fig:monojet comparison}
\end{center}
\end{figure*}

Our first observation concerns the importance of including events with a second hard jet in the merged sample. Naively one would expect that in monojet searches the MET distribution (used to select the various signal regions) would be fairly insensitive to the exact $p_T$ distribution of the second jet. In fact, the analyses of both collaborations only impose a veto on the third jet\footnote{In~\cite{CMS-PAS-EXO-12-048,ATLAS:2012zim,ATLAS:2012zim} events with more than two jets with $p_T$ above 30 GeV in the region $|\eta| < 4.5$ are rejected. }, together with a restriction on angular distribution of the second jet. This implies that events with two hard jets are likely to pass monojet cuts and must therefore be included in the simulation to properly reproduce the missing $E_{T}$ distribution. Indeed, the bounds one would obtain not including a second additional jet in the merged sample are dramatically weaker. The difference is shown in Fig.~\ref{fig:monojet comparison}, where we compare the excluded couplings from the CMS analysis~\cite{CMS-PAS-EXO-12-048}  for the sample  $pp \rightarrow \chi\bar\chi+(0,1)j$ (dashed line) and  $pp \rightarrow \chi\bar\chi+(0,1,2)j$ (dotted line). 

At this point one should should question the validity of simulating $pp \rightarrow \chi\bar\chi+(0,1,2)j$, given the fact that the model contains colored massive resonances decaying into light partons, which can potentially go on-shell: $pp \rightarrow \chi\bar\chi+(0,1,2)j$ contains processes of formally different order in the $\alpha_{s}$ expansion, such as $pp \rightarrow \tilde q\tilde q, \tilde q \rightarrow \chi j$ or $pp \rightarrow \bar\chi \tilde q + (0,1)j,  \tilde q \rightarrow \chi j$. These processes produce extra jets in the squark decays which should not be matched and have different hard scales and may require different matching scales. Moreover requiring ${pp \rightarrow \chi\bar\chi+(0,1,2)j}$ would not include any extra initial state radiation (ISR) jets for the case of squark pair production. This is a problem when the splitting between the mass of the mediator and the DM mass decreases, since the jets generated in the hard matrix elements coming from the squark decay would become very soft.
 
The standard procedure to deal with these mass ``singularities'' is to subtract the on-shell contribution from the regularized Breit-Wigner resonance propagator and to generate different samples of events with the heavy resonance in the final state~\cite{Beenakker:1996ch}, together with the needed number of hard jets.  One then merges the sample and makes the unstable states decay. MadGraph allows this subtraction by vetoing  the phase space integration in the neighborhood of any internal resonance mass-shell, since those regions of phase space are populated by the resonant on-shell decay. Numerically, the neighborhood of the squark mass-shell is defined to be much wider than the resonance width, to guarantee the correct factorization of rates in terms of cross-sections and branching ratios. 

Following this procedure, we split events with two, one and zero on-shell squarks in the final state and separately merged these samples. Double counting of events is avoided by the on-shell resonance subtraction procedure described above. Schematically we generated:
\begin{enumerate}
\item $p p \rightarrow \tilde{q} \tilde{q}^\dagger + (0,1,2)j$;
\item $p p \rightarrow \chi \tilde{q}^\dagger, \bar\chi \tilde{q}  + (0,1,2)j$ with no mass-shell integration for internal squark lines;
\item $p p \rightarrow \chi \bar\chi  + (0,1,2)j$ with no mass-shell integration for internal squark lines.
\end{enumerate}
In the following we will refer to the combination of the above three samples as the \emph{split sample}.

This method represents a valid technique whenever the resonance width is sufficiently small to justify a narrow width approximation (NWA). Unfortunately, in the analysis we are carrying out, this assumption does not appear well founded over the whole parameter space we will consider. 
 In fact, as we will see in the next subsections, monojet and jets+MET searches are sensitive only to $O(1)$ couplings, which translates to relatively large widths.

In fact if $\Gamma_{M}$ becomes larger, then the veto window increases to the point where very little phase space is left for process (3). Moreover, for computational speed limitations we perform the squark decays in Pythia, which forces them to decay exactly on shell, further introducing kinematical distortion when $\Gamma_{M}$ is large. Therefore we are forced to constrain the veto window to smaller values, of the order of $\Gamma_{M}$, thus contaminating off-shell mediator samples with quasi-on-shell events and breaking the $\sigma \cdot BR$ NWA factorization formula. We account for these effects by properly rescaling the samples (1) and (2) before recombining the events.  Rescaling of the samples (1) and (2) introduces an additional uncertainty for large widths, for which we provide additional details in Appendix~\ref{appendix:Analysis}.  Note that this method also does not capture the interference between different samples around the squark mass-shell which, while totally negligible in the NWA, may become progressively more important as the squarks become broader.

The results of this split sample procedure are shown in Fig.~\ref{fig:monojet comparison} as the continuous line.  Comparing the $pp \rightarrow \chi \chi + (0,1,2)j$ and split samples, we see that they give comparable limits when $m_M - m_{DM} \gtrsim 300 $ GeV.  The reason is that in this region the hard matrix element already contains one or two hard jets which are needed to pass the analysis cuts, while all the soft radiation is well described by the shower. Moreover, given the relatively large DM coupling and squark width, the usual power counting issues associated with internal colored resonances going on-shell are less problematic. In the more compressed region, however, when $m_M - m_{DM} \gtrsim 300 $ GeV, additional jets in the matrix element must be included to properly account for ISR.  This is resolved by adding additional jets alongside the squark production ({\em e.g.} $pp\rightarrow \tilde q \tilde q + (0,1,2)j$ and $pp\rightarrow \tilde q \chi + (0,1,2)j$). In this mass range, the presence of additional hard ISR is fundamental for passing the $p_{T}$ cut on the leading jet and to generate enough missing $E_{T}$, since processes like the ones shown in Fig.~\ref{fig:feynmanDM}, panel (b,d), or Fig.~\ref{fig:feynman_2j}, panel (a,b), would be completely neglected due to the softness of the emitted jets.  The result is a change in the constraint on $g_M$ as large as 50\% in the compressed region.

\begin{figure*}[htb]
\noindent\makebox[\textwidth]{
\centering
\includegraphics[scale=0.45]{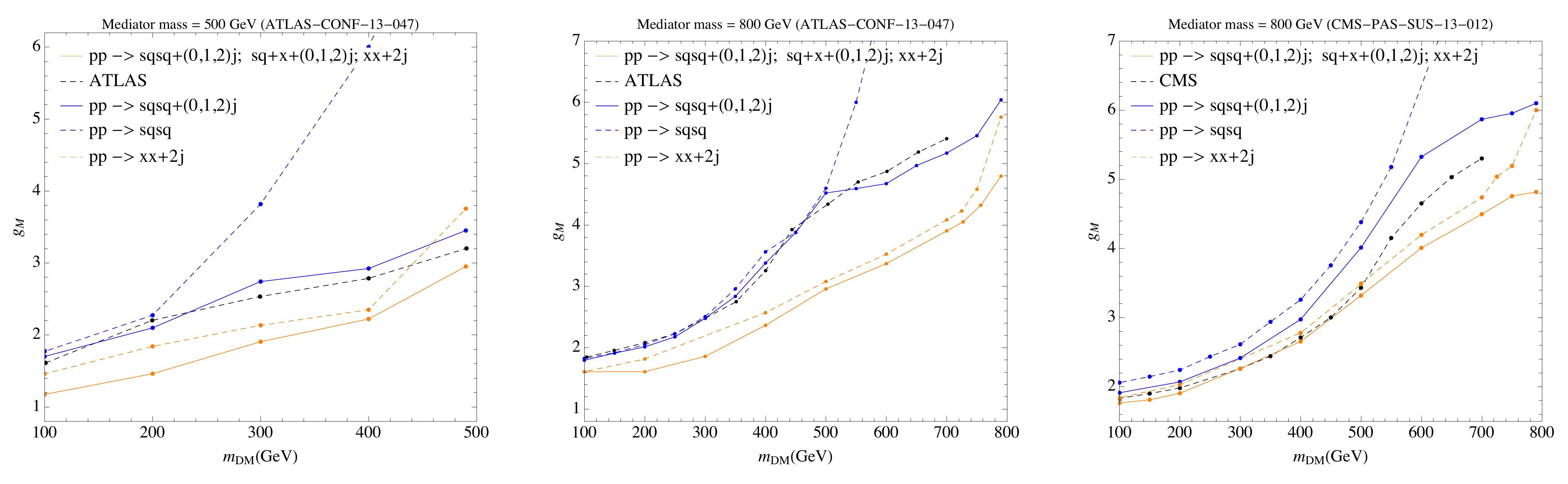}}
\caption{ Exclusion limits from jets+MET searches for mediator masses $m_M = 500,~800$ GeV. The black dashed line corresponds to limits on cross-section by the collaborations ((a,b) ATLAS, (c) CMS); the dashed blue represents squark on-shell production with no additional jets; the continuous blue represents squark on-shell production with up to two additional jets, and as a quality check we can see that this result agrees quite well with the results of ATLAS which assumes only on shell squark production when extracting their constraint (our constraint is weaker than CMS for reasons we describe in the text).  Including both on and off-shell squark production strengthens the bounds and is shown by the orange lines.  The orange dotted line represents events with two jets and met with no further constraints, while the continuous orange line is the combination of three samples as described in Sec.~\ref{sec:mono}.  See the text for further details.}
\label{fig:dijets comparison}

\end{figure*}

\subsection{jets+MET: off-shell production}\label{sec:dijets}

We now move on to the jets+MET analysis.  Our constraints on $g_M$ from the jets+MET analyses of ATLAS (panels (a) and (b)) and CMS (panel (c)) are shown in Fig.~\ref{fig:dijets comparison}.  Similar to the monojet constraints, obtaining the correct limit outside the NWA requires that we consider multiple samples in our analysis, which we describe here.  Among the signal regions giving rise to the strongest constraints, both CMS and ATLAS require two and only two hard jets, so that one would expect that producing  $pp \rightarrow \chi\bar\chi j j$ would capture the bulk of the limit when $\Gamma_{M}$ is not small.  We would like to stress that we do not impose any restriction on the intermediate states.  
In this we differ substantially from~\cite{An:2013xka,Chang:2013oia,Bai:2013iqa}. The procedure taken in those works is to generate events for on-shell production of squarks, which later decay into quarks and DM, and to compare the obtained cross-sections with the simplified models results provided by ATLAS and CMS.  The results of such a procedure, with constraints as extracted by the collaborations, are shown as the black dashed line in Fig.~\ref{fig:dijets comparison}.  The squark-neutralino simplified model implicitly assumes that the cross-section is dominated by diagrams with mostly on-shell squarks and that their width is extremely narrow.  On the other hand, the values of the coupling to which jets+MET searches are sensitive force the squark widths to be comparable or larger than the $p_{T}$ thresholds of the jets required by the analyses so that finite width effects are important.

Rather than simply applying the results of the collaborations using the constrained cross-sections that they quote, we re-compute the efficiencies by simulating events through to the DM and jet final states and applying the detector simulation, as described in Sec.~\ref{sec:mono}.  We begin with a validation of our results by simulating the merged sample $p p \rightarrow  \tilde{q}\tilde{q}^\dagger +(0,1,2)j $ in the NWA and comparing it to the results extracted with the squark-neutralino simplified model. The results for our simulation are shown in Fig.~\ref{fig:dijets comparison} as blue lines, compared to the results of the collaborations with the dashed black line.  Note that, in order to reproduce the simplified model efficiencies of the collaborations,  the presence of additional jets is crucial for passing the hard cuts on the jet $p_{T}$ already in a moderately squeezed region: the dotted blue line shows how a sample of events for  $p p \rightarrow \tilde{q} \tilde{q}^\dagger$ would completely fail for $m_M - m_{DM} \leq 300 $ GeV,  while at small DM masses there is practically no difference.  
The agreement with ATLAS is quite good  and implies that the differences between various methods are not due to the processing of the events.  As can be seen in the right panel of Fig.~\ref{fig:dijets comparison}, however, in the case of a CMS analysis, while the shape of the exclusion contour is fairly well reproduced, we were unable reproduce its normalization. This is not surprising since Refs.~\cite{CMS-PAS-SUS-13-012,CMS:zxa,CMS:2012kba,Chatrchyan:2012lia}  combine the various signal regions in a multi-bin likelihood without providing sufficient information regarding the bin-to-bin correlations for the SM background expectation. Our curve shown in Fig.~\ref{fig:dijets comparison} is based on a multi-bin fit neglecting background correlations among the various bins. Given the fact that the Atom analysis framework reproduces the efficiencies and the signal distributions provided in the experimental papers within $10-15\%$, we believe that this discrepancy is entirely due to the differences in the statistical analyses. Setting the limit with the $CL_s$ statistical analysis described in Appendix~\ref{appendix:CLS}, our bounds are constantly more conservative than those provided by the collaboration. 

After validating the simulation procedure, we seek to quantify the error made in assuming a NWA, as utilized in~\cite{An:2013xka,Chang:2013oia,Bai:2013iqa}.  In Fig.~\ref{fig:dijets comparison}, as the dashed orange line, we show exclusion bounds obtained simulating the production of $p p \rightarrow jj\chi \bar\chi $, and we can see that they differ quite substantially from the NWA, the latter being less constraining almost everywhere. It is worth noticing that the main sources of discrepancy between the dashed orange and the continuous blue line are the finite-width effects for the resonant squarks and the presence of additional diagrams where the DM is produced in association with a squark, effects that are not captured by the blue line.  As expected, the sample $p p \rightarrow jj\chi \bar\chi $ fails to describe correctly the model in the compressed region, for the same reasons discussed in Sec.~\ref{sec:mono}, and we need to use the split sample of events generated according to the discussion there.  This is shown as a continuous orange line in Fig.~\ref{fig:dijets comparison}. Note that, contrary to monojet bounds, in SUSY multi-jet searches extra ISR jets are important only for  $m_M - m_{DM} \leq 100$ GeV.  

To summarize, the blue line and the dotted orange line provide a completely self-consistent comparison of the effects missed by restricting to the NWA. Our conclusion is that this assumption greatly weakens the exclusion bounds, changing the result by an $O(1)$ factor in a sizable part of the parameter space.

\section{\textbf{Monojet versus Dijet Searches}}
\label{sec:monojet versus dijet}

\subsection{Results}

Having validated our results and established our method in the previous section, in this section we will provide a complete scan of results in the $m_{DM}-m_M$ plane, extracting a constraint on the effective EFT scale $\Lambda \equiv m_M/g_M$, that can be used for translating our results to the DD plane.   We present the results of our analysis for the case of $\tilde{u},\, \tilde{d},\,\tilde{c},\,\tilde{s}$, $L+R$ and the other extreme case of only two squarks $\tilde{d}_R,\, \tilde{s}_R$. 
For each pair $m_{DM}$, $m_M$, we present here only the strongest bound obtained among all the searches from CMS~\cite{CMS-PAS-SUS-13-012,CMS:zxa,CMS:2012kba,Chatrchyan:2012lia,CMS-PAS-EXO-12-048,Chatrchyan:2012me} and ATLAS~\cite{TheATLAScollaboration:2013fha,ATLAS:2012ona,ATLAS-CONF-2012-033,TheATLAScollaboration:2013aia,ATLAS:2012zim,ATLAS:2012ky}. It is worth mentioning that, though our CMS results are conservative on account of the statistics (see Sec.~\ref{sec:dijets}) they represent our strongest constraint. This is not surprising since the combination of the various signal region bins provide more statistical power than the single-bin exclusion performed by other analyses. We would therefore expect even stronger bounds from jets+MET if the statistical details of the analysis were available.

\begin{figure*}[t]
\begin{center}
\subfigure[]{\includegraphics[scale=0.58]{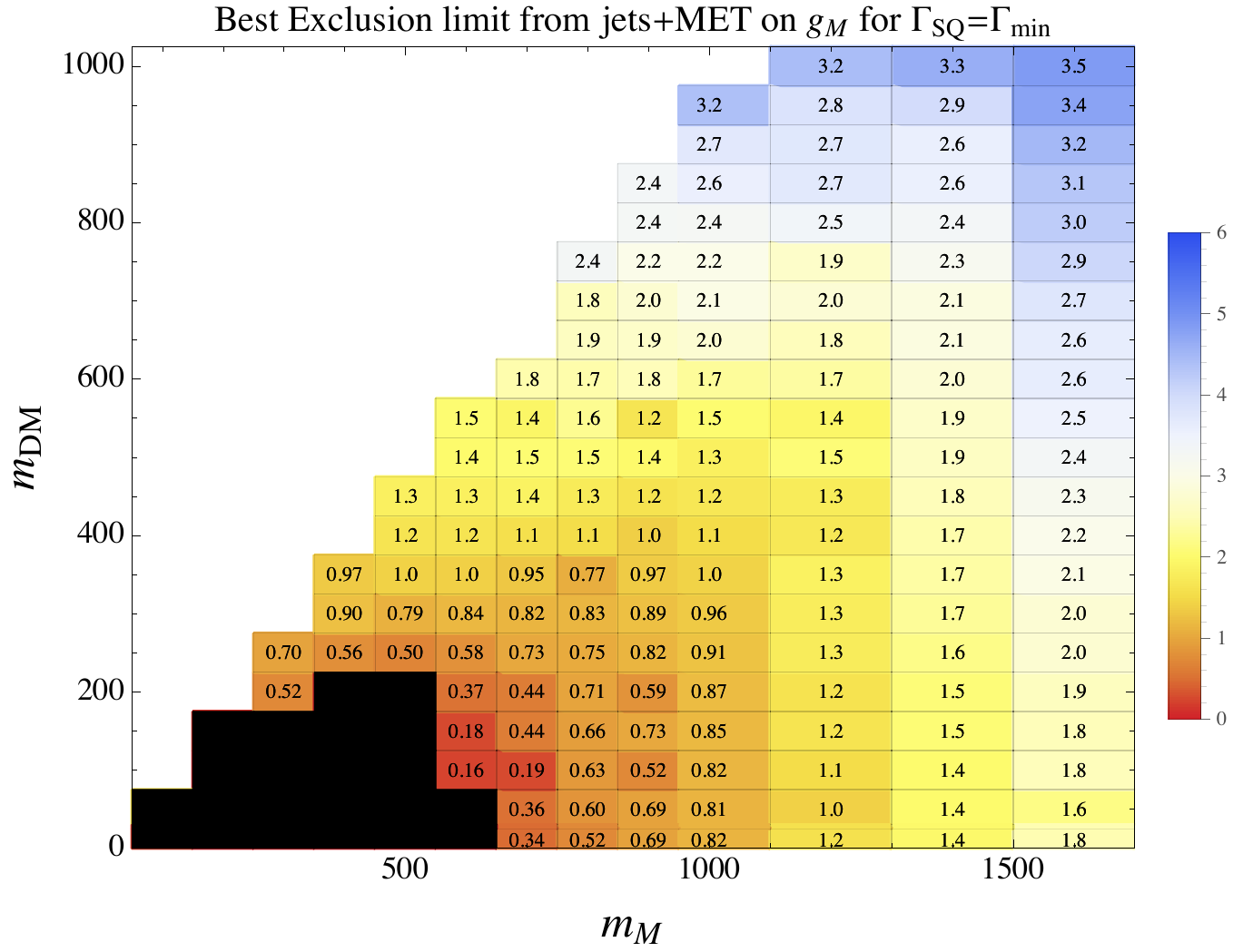}}
\subfigure[]{\includegraphics[scale=0.58]{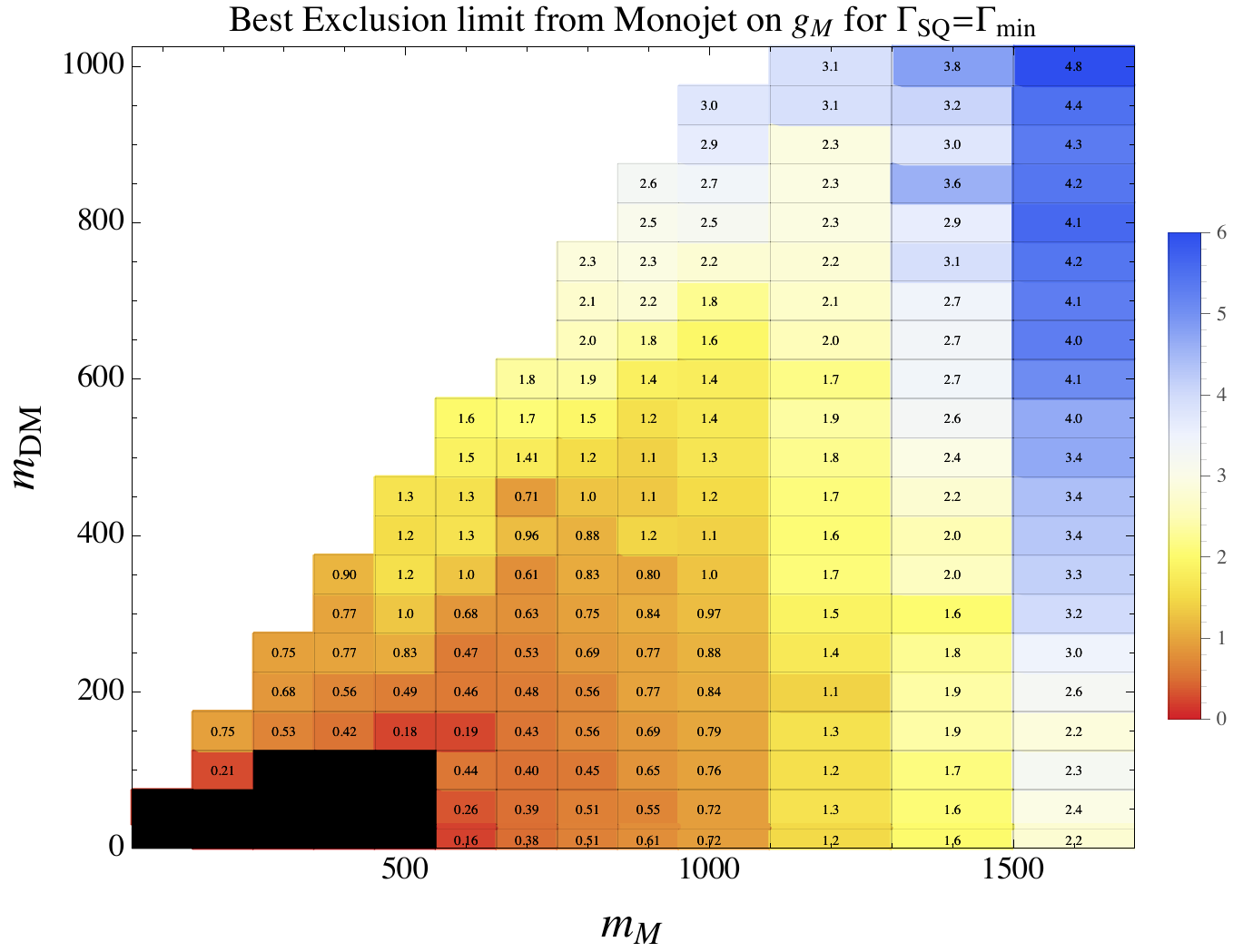}}
\caption{Limits on $g_{M}$ (for the case of mediator coupling to $\tilde{u},\, \tilde{d},\,\tilde{c},\,\tilde{s}$, $L+R$) from ({\em left}) jets+MET, and ({\em right}) monojet, for a mediator decaying only to DM and a quark, with the natural width computed from Eq.~\ref{Gammamin}. The black region in (a) is excluded from the pure QCD production of the mediator.}
\label{fig:t-channel}
\end{center}
\end{figure*}

\begin{figure*}[t]
\noindent\makebox[\textwidth]{
\centering
\includegraphics[scale=0.55]{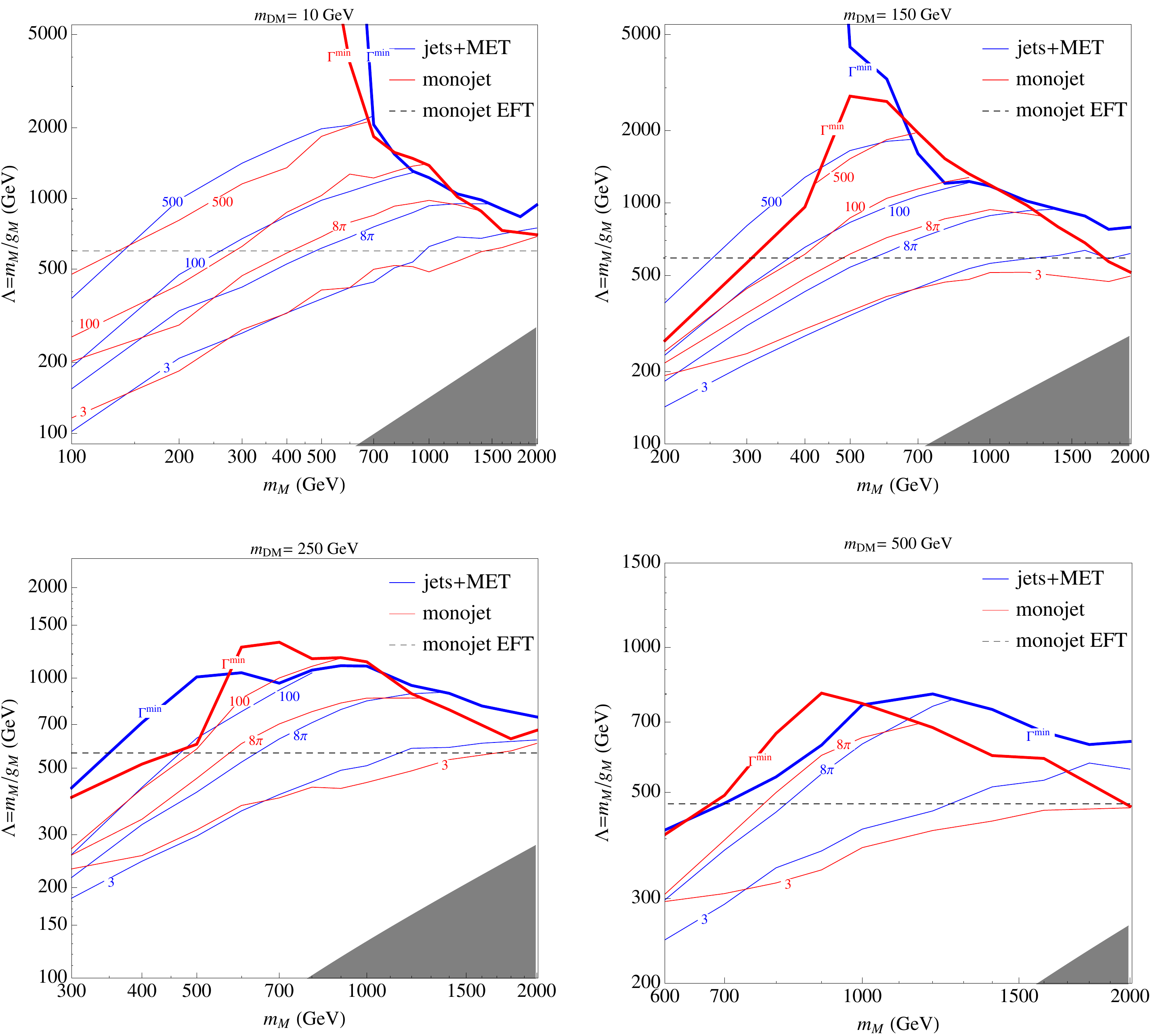}}
\caption{Constraint on $\Lambda = m_M/g_M$ from monojet (red curves) and jets+MET (blue curves) for various choices of the mediator width.  The labels on each line correspond to the width, written as $m_M/\Gamma_M$, and in addition we show the constraint from the natural width,  $\Gamma_M^{min}$, in Eq.~\ref{Gammamin}.    A line will stop when it merges with the $\Gamma_M^{min}$ line, since we require the width of the particle to be larger than its natural width to decay to a quark and the DM, $\Gamma_M > \Gamma_M^{min}$. The dashed black line represents the EFT limit. The grey shaded region is where the natural width of the mediator would be larger than its mass.}
\label{fig:t-channel Lambda}
\end{figure*}

We begin with the case of $\tilde{u},\, \tilde{d},\,\tilde{c},\,\tilde{s}$, $L+R$.  The constraints on the coupling $g_M$ are shown in Fig.~\ref{fig:t-channel}.  The first important remark we make is that the jets+MET search is able to exclude the model with $m_{M}\lesssim 600$ GeV and $m_{\chi}\lesssim 250$ GeV, when the width of the mediator is set by the natural decay width to the DM and a quark.\footnote{The width can of course be larger if additional particles contribute to the decay.}  This is because, at small enough coupling $g_M$, the production is dominated by the (irreducible) QCD production initiated by two gluons.  $g_M$ is also small enough for the mediator width to be narrow, so that neither production nor decay depends on $g_M$.  In this case, the model is excluded for all $g_M$ and hence all $\Lambda \equiv m_{M}/g_M$. 

As can be seen in the right panel of Fig.~\ref{fig:t-channel}, the same is true for the monojet case, although the excluded region is smaller and limited to  $m_{M}\lesssim 500$ GeV and $m_{\chi}\lesssim 150$. It must be stressed that this represents a novelty of the 8 TeV analysis which does not impose any cut on the second jet transverse momentum, rendering the monojet searches more similar to a standard squark search (with the angular cuts and the MET requirements forcing $m_{eff}$ to be sizable as well). On the contrary previous searches only selected processes with a single hard jet, essentially restricting to the diagrams of Fig.~\ref{fig:feynmanDM}. In the presence of those cuts the cross-section decreases monotonically with $g_{M}$, since all the processes irreducibly depend on $g_M$. Hence 7 TeV-like monojet searches are not able to exclude completely the model.
In general one finds that the bounds from SUSY searches tend to be slightly stronger than those from monojet searches, with the latter becoming comparable or slightly stronger (by only ${\cal O} (10\%)$ on $g_{DM}$), for mediator masses around $1\,{\rm TeV}$ and very light DM, which is probably beyond what our recasting and statistical procedure can resolve.

\begin{figure*}[htb]
\noindent\makebox[\textwidth]{
\centering
\includegraphics[scale=0.55]{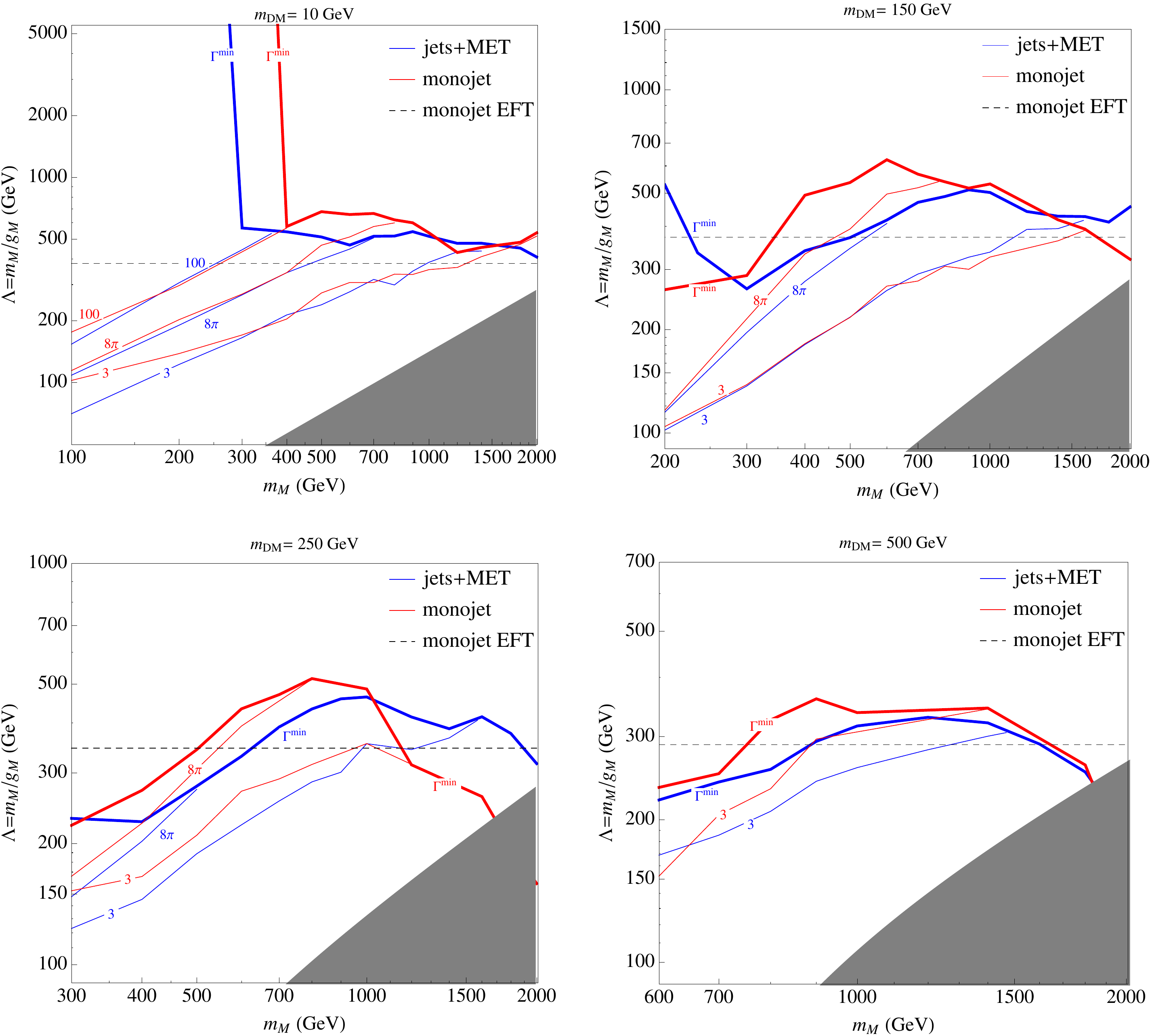}}
\caption{Exclusion bounds for a DM model containing only squarks $\tilde{d}_R$ and $\tilde{s}_R$ .  Conventions are as in Fig.~\ref{fig:t-channel Lambda}.}
\label{fig:t-channel Lambda drsr}
\end{figure*}

 In Figs.~\ref{fig:t-channel Lambda},~\ref{fig:t-channel Lambda drsr}, for the cases of $\tilde{u},\, \tilde{d},\,\tilde{c},\,\tilde{s}$ $L+R$ and $\tilde{d}_R,~\tilde{s}_R$ respectively, we present our results as lower bounds on the scale $\Lambda=m_{M}/g_M$.  The results depend strongly on the width of the squark, so we choose the natural width $\Gamma_M=\Gamma_M^{min}$ as well as larger widths $\Gamma_M = m_{M}/500,\, m_{M}/100,\, m_M/(8 \pi)$ and $m_M/3$.   
We compare the lower limits on $\Lambda$ from jets+MET (blue) and monojet (red) as a function of $m_M$ for several values of $m_{DM}$.  As expected, the strongest constraint is derived from the narrowest width mediator.  Moreover, the exclusion limit, for any $g_{M}$ when $m_{M}$ is sufficiently light, that was found for $\Gamma_{M}=\Gamma_{M}^{\rm min}$ disappears when restricting to fixed widths. This is due to a branching ratio suppression $\Gamma_{M}^{\rm min}/\Gamma_{M}$ for squark decays.

 In each plot, as a dashed line, we also show for comparison the bound on the scale $\Lambda$ arising from assuming a valid EFT.\footnote{In practice we obtained the limit setting the mediator mass to 5 TeV and running the same analysis as for the other case.}  For masses larger than 2 TeV all the limits converge towards the EFT line and the dependence on the mediator width becomes negligible.

 The first message from these figures is that there is no such a thing as a lower bound on $\Lambda$; instead, there are bands, parameterized by the details of the theory. Secondly, the relative strength of monojet and jets+MET constraints varies with the masses. For light mediators and natural widths, jets+MET searches largely dominate over the rest. For larger widths, however, we observe very small differences and the limits become comparable.

In these plots we can also observe the failure of the EFT approach. The black dashed line shows the exclusion limit one would obtain using an effective four fermion interaction (here mimicked by a mediator with a mass at 5 TeV). On the other hand, relatively light squarks can be produced on shell, so that the jets+MET search provides a powerful direct probe. The above is true only for the natural width case: a larger width always gives bounds weaker than the EFT bounds.
As we increase the mediator mass both monojet and jets+MET bounds relax. If we were to extend our analysis to squark masses of a few TeV we would observe the bound converging to the EFT bound, though by the time this happens the perturbative interpretation of the mediator as an elementary scalar meditating a tree level interaction between SM and DM sector is lost. In fact, as is clear from Figs.~\ref{fig:t-channel Lambda},~\ref{fig:t-channel Lambda drsr}, the grey region (namely when $\Gamma_M > m_M$) will extend above the bound at $m_M \sim 2-3$ TeV.

\begin{figure*}[htb]
\noindent\makebox[\textwidth]{
\centering
\includegraphics[scale=0.45]{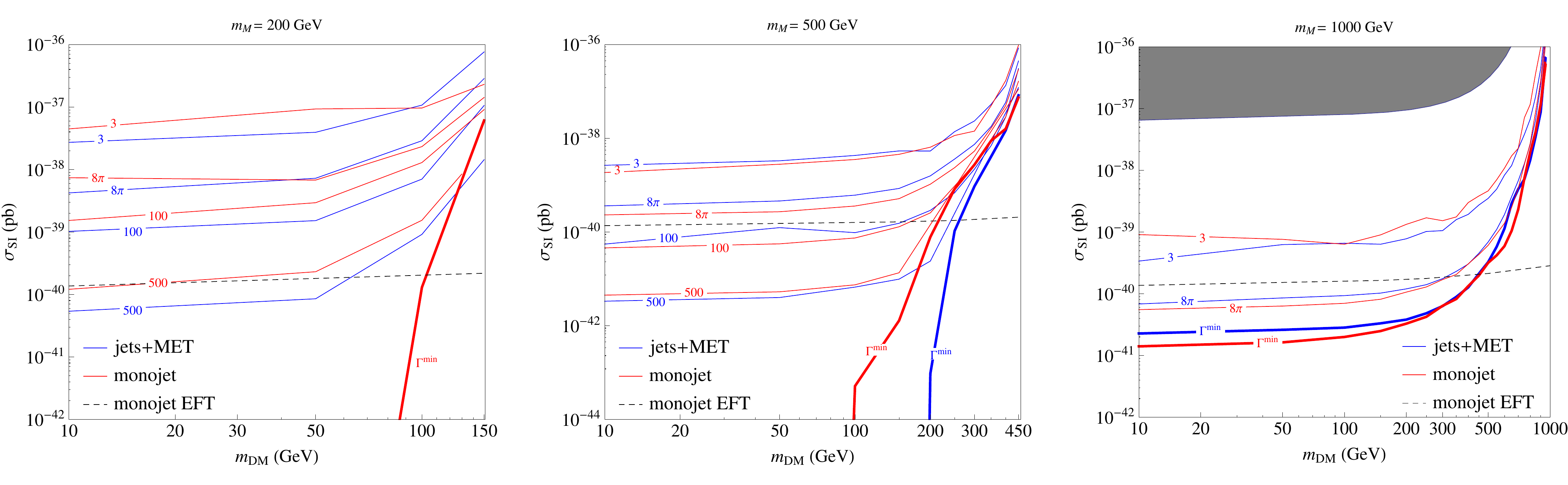}}
\caption{Monojet (in red) and jets+MET (in blue) bounds on the DD cross-section as a function of the DM mass at fixed mediator mass.  The labels on each line correspond to the width, written as $m_M/\Gamma_M$.  The grey region corresponds to the particle becoming very broad, $\Gamma_M^{min} \geq M$, so that the perturbative approach we apply is invalid.  In the left panel no blue line appears because the whole region of parameter space is ruled out.}
\label{fig:t-channel sigma}
\end{figure*}

\begin{figure*}[htb]
\noindent\makebox[\textwidth]{
\centering
\includegraphics[scale=0.45]{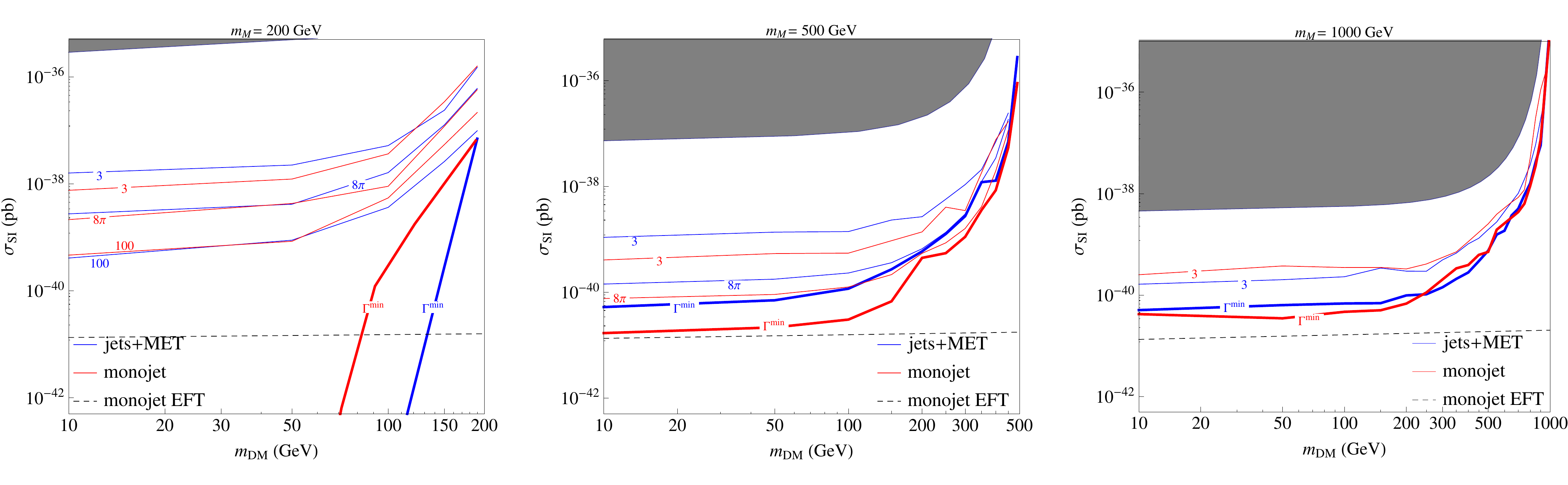}}

\caption{Exclusion bounds for a DM model containing only squarks $\tilde{d}_R$ and $\tilde{s}_R$ .  Conventions are as in Fig~\ref{fig:t-channel sigma}. }
\label{fig:t-channel sigma drsr}
\end{figure*}

Finally, in Figs.~\ref{fig:t-channel sigma},~\ref{fig:t-channel sigma drsr} we translate these bounds  in the $m_{DM}-\sigma_{n}$ plane, again for the cases of $\tilde{u},\, \tilde{d},\,\tilde{c},\,\tilde{s}$ $L+R$ and $\tilde{d}_R,~\tilde{s}_R$ respectively with the same choices for the width as in Fig.s~\ref{fig:t-channel Lambda},~\ref{fig:t-channel Lambda drsr}.  For a Dirac particle there is both spin-independent and spin-dependent scattering, though the dominant process will be spin-independent, for which the formula is
\begin{equation}
\sigma_{SI} = \frac{\mu_r^2}{64 \pi} \frac{1}{|m_M^2-m_{DM}^2- i \Gamma_M m_M|^2} \frac{(Z f_p + (A-Z) f_n)^2}{(Z + (A-Z))^2},
\end{equation}
where $\mu_r$ is the nucleon-DM reduced mass, and 
\begin{equation}
f_p = 2 {g_{L,M}^u}^2+{g_{L,M}^d}^2+2 {g_{R,M}^u}^2+{g_{R,M}^d}^2,~~~~~f_n = {g_{L,M}^u}^2+2{g_{L,M}^d}^2+{g_{R,M}^u}^2+2{g_{R,M}^d}^2,
\end{equation}
with $ g_{L,M}^u,~g_{L,M}^d,~g_{R,M}^u,~g_{R,M}^d$ the coupling of the mediator $M$ to left or right handed up or down quarks.


\subsection{Compressed spectrum}

\begin{figure*}[htb]
\noindent\makebox[\textwidth]{
\centering
\subfigure[]{\includegraphics[scale=0.9]{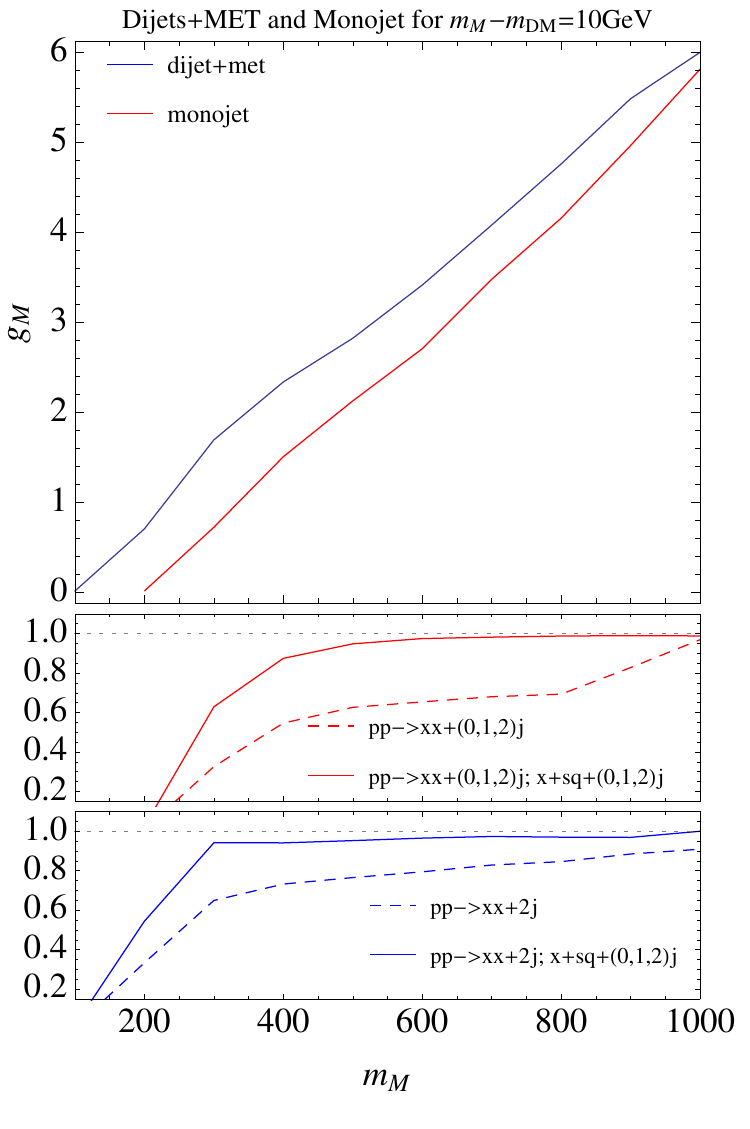}}
\subfigure[]{\includegraphics[scale=0.9]{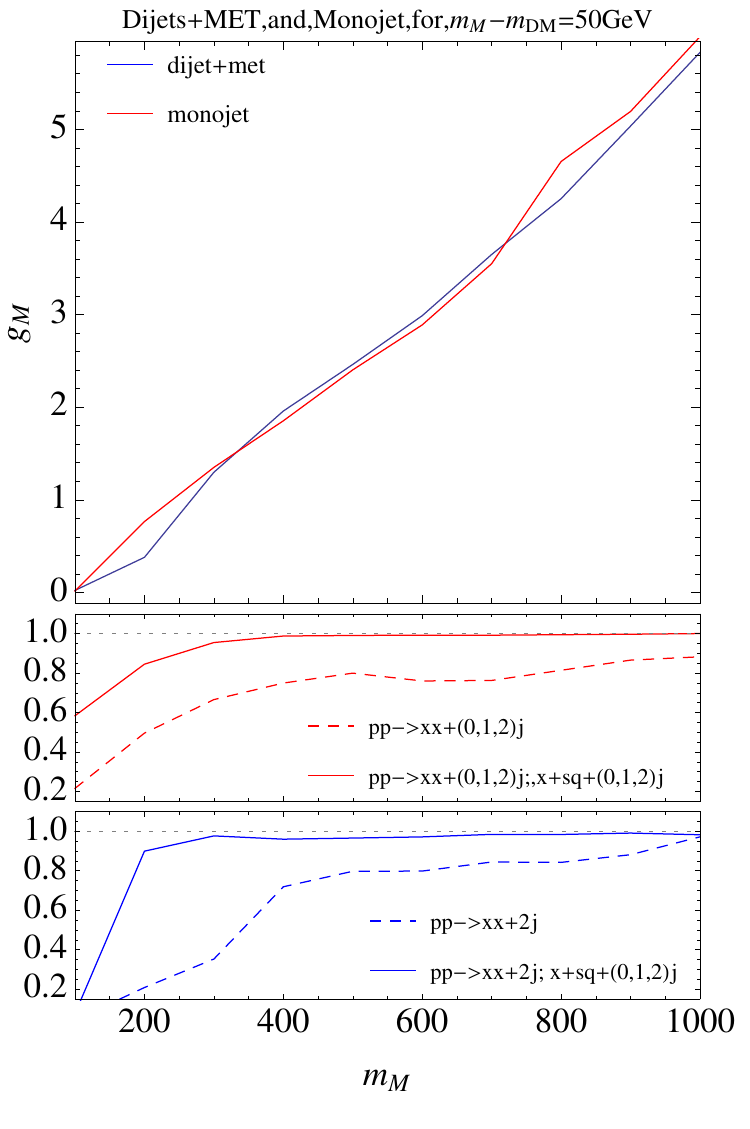}}
\subfigure[]{\includegraphics[scale=0.9]{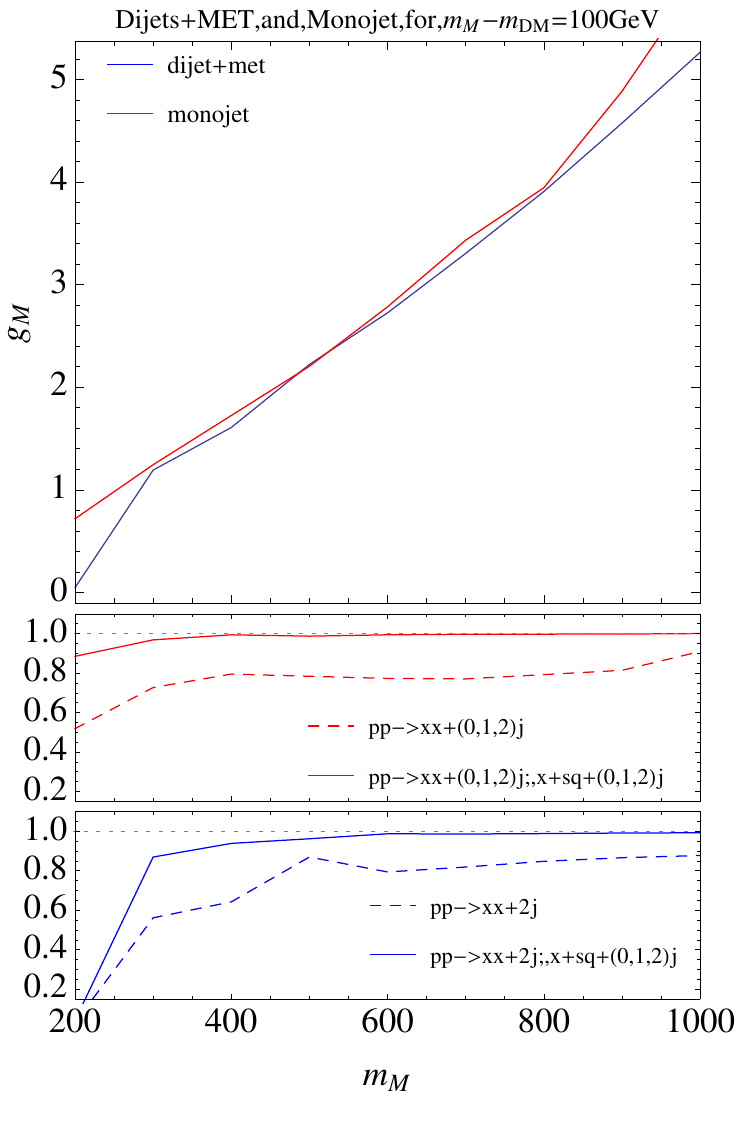}}
}
\caption{Constraints from dijet+MET (blue) and monojet (red) on $g_M$ in the compressed region, with $m_M-m_{DM} = 10,~50,~100$ GeV.  The constraints from dijet and monojet are competitive. On the bottom: ratio of sub-samples over the complete sample for monojet (red) and jets+MET(blue)}  
\label{fig:compressed}

\end{figure*}

Lastly, in Fig.~\ref{fig:compressed}, we focus on the case of the compressed spectrum, presenting the results for the model containing only $\tilde{d}_R,~\tilde{s}_R$ and mass splittings $m_M-m_{DM} = 10,~50,~100$ GeV.  In the upper panels we show the exclusion curves both for monojets and jets+MET searches, while in the lower panels we present the decomposition of the limits in terms of the three event samples introduced in Sect.~\ref{sec:mono}, defined as the ratio of excluded couplings from the partial to the full set of samples. As one can see the dominant contributions in both cases are coming from DM pair production and DM squark associated production, with ISR fulfilling the jet cuts requirements of both analyses. The rate is therefore controlled by $g_{M}$ and increases monotonically with the mediator mass. Pair production of the mediator is only relevant at low mediator masses when the constrained coupling is dropping below 1 and the pure QCD production mechanism takes over. While one might expect that the dijet constraints would be stronger than monojets, we find that jets+MET is competitive with monojets, except for the extremely degenerate case $\Delta M \sim 10\, {\rm GeV}$. 

%

\section{\textbf{Conclusions}}\label{Sec:Conclusions}

The goal of this paper was to compare monojet against jets+MET bounds on DM production at the LHC for the case of $t$-channel mediators, such as a squark. The jets+MET channel constrains direct production of the mediator, which is important when outside the regime of validity of the EFT.  We examined the constraints in all regions of parameters space, both when the EFT is valid and outside the regime of validity of the EFT.  Unlike previous works, we included all the relevant processes leading to the production of one or more jets and missing $E_{T}$ both for the case of monojet searches and for SUSY jets+MET ones, without limiting to those processes that are dominant in the narrow width for the mediator particle; we found that including these additional processes leads to an ${\cal O}(1)$ change in the constrained coupling.  Especially when the mediator is light, the jets+MET constraint usually dominates over the monojet constraint; at heavier mediator masses the monojet and jets+MET constraints are competitive, with monojet searches signifcantly outperforming jets+MET searches only in the extremely degenerate region (mass splittings of the order of $10\,{\rm GeV}$ of parameter space. 

It is clear from our study that when applying monojet constraints to a model of DM, specifying the UV completion is essential both to correctly assessing the validity of the monojet constraints as well as to considering the presence of other important processes that yield stronger constraints. Therefore, Dark Matter collider searches (contrary to Direct and Indirect searches) are better discussed in a simplified model language than in the language of operators in an EFT. In our next paper, we will turn to considering these effects on the other class of simplified models, based on  $s$-channel mediators.

%
%

\section*{\textbf{Acknowledgments}}
We would like to thank S. Alioli for helpful discussions. This research used resources of the National Energy Research Scientific Computing Center, which is supported, as the work of AV, by the Office of Science of the U.S. Department of Energy under Contract No. DE- AC02-05CH1123. KZ is supported by NSF CAREER award PHY 1049896. MP would like to thank the CERN TH group for its warm hospitality, and MP and KZ would like to thank the Aspen Center for Physics where part of this work was completed.

%
%

\appendix

\section{\textbf{Simulation Details}}\label{appendix:simulation}

In the present section we provide details of simulation and combination of events, explicitly including the process generation commands run on MadGraph5~\cite{Alwall:2011uj}. For concreteness, we take a model with two squarks, labeled \texttt{dr},\texttt{sr} and a DM Dirac fermion labeled \texttt{chi}.

\addtocontents{toc}{\protect\setcounter{tocdepth}{1}}
\subsection{Dijet sample}
\addtocontents{toc}{\protect\setcounter{tocdepth}{2}}

In order to compute exclusion bounds from the jets+MET analysis we generated the following three samples for a given pair of $m_{DM}, m_M$
\begin{verbatim}
generate p p > j j chi chi~ / a h z w+ w- QED=n QCD=m
\end{verbatim}
where $(n,m)={ (4,2),(4,0),(2,2)}$, with the QED value representing here the maximum number of powers of $g_{DM}$ in a Feynman diagram.\footnote{Therefore the first sample contains the other two and the interference.}  For each sample we generated events with different values of the ratio of $\Gamma_M/m_M$. Note that the samples with $(n,m)={(4,0),(2,2)}$  scale homogeneously with the coupling $g_{DM}$. Calling $g_{DM}^{sim}$ the actual value of the coupling used to generate the events, one can compute the cross section times the efficiency in the i-th bin as a function of  $g_{DM}$ as
\bea
n_i (g_{DM})= \sigma_{4,0}\, \varepsilon_{4,0}^i \left(\frac{g_{DM}}{g_{DM}^{sim}}\right)^8 + \sigma_{2,2} \,\varepsilon_{2,2}^i \left(\frac{g_{DM}}{g_{DM}^{sim}}\right)^4 + \left( \sigma_{4,0}\, \varepsilon_{4,2}^i- \sigma_{4,2} \,\varepsilon_{4,0}^i- \sigma_{2,2} \,\varepsilon_{2,2}^i\right) \left(\frac{g_{DM}}{g_{DM}^{sim}}\right)^6,\nonumber\\
\eea
where $ \sigma_{n,m}$ and $\varepsilon_{n,m}^i$ represent the cross section and i-th bin efficiency for the sample $(n,m)$. For any choice of the mediator width $\Gamma$ one can compute the maximal value allowed given the experimental observation, under the constraint that $g_{DM}$ cannot exceed the the value $g_{DM}^{max}$ imposed by the width condition $\Gamma_{M}^{min}(g_{DM}^{max}) = \Gamma$, with $\Gamma_{M}^{min}$ defined in Eq.~\ref{Gammamin}. Generating samples for different values of  $\Gamma/m_M$ we obtained an interpolation grid that can be used to extract the bounds shown in the main text.   

\subsection{Monojet sample}

The method used for the monojet analysis follows closely the discussion of the previous subsection. The only difference consists in the presence of events including 0 and 1 jet only:

\begin{verbatim}
generate p p >  chi chi~ / a h z w+ w- @1
add process p p > j  chi chi~ / a h z w+ w- @2
add prcess p p > j j chi chi~ / a h z w+ w- QED=n QCD=m @3,
\end{verbatim}
where $(n_1,n_2)={ (4,2),(4,0),(2,2)}$.

\subsection{Split sample}

As discussed in Sec.~\ref{sec:mono} in order to access the compressed spectrum region we need a method that correctly merges higher jet multiplicity events and at the same time does not rely on the NWA. We achieved this by producing and combining appropriately three event samples. The first sample consists of squark pair production plus up to 2 jets. Notice that we do not restrict to QCD induced jets only:  
\begin{verbatim}
define sq = dr sr dr~ sr~
define dm = chi chi~
generate p p >  sq sq / a h z w+ w- QED=2 QCD=2 @1
add process p p > j  sq sq / a h z w+ w- QED=2 QCD=3 @2
add prcess p p > j j sq sq / a h z w+ w- QED=4 QCD=4 @3.
\end{verbatim}
The second sample contains events with squark-DM associated production plus additional jets. In order to not double count events we vetoed a second  squark on shell:
\begin{verbatim}
generate p p >  dm sq / a h z w+ w- $sq QED=1 QCD=1 @1
add process p p > j  dm sq / a h z w+ w- $sq QED=1 QCD=2 @2
add prcess p p > j j  dm sq / a h z w+ w- $sq QED=3 QCD=3 @3.
\end{verbatim}
Finally we must include all the events with two hard jets where there are and no squarks on shell:
\begin{verbatim}
generate p p >  chi chi~ / a h z w+ w- $sq @1
add process p p > j  chi chi~ / a h z w+ w- $sq @2
add prcess p p > j j chi chi~ / a h z w+ w- $sq QED=4 QCD=2 @3.
\end{verbatim}
Splitting further the samples into $g_{DM}$ homogenous  sub-samples to be able to rescale the cross sections would be highly time and computing costly and we did not do it, instead creating grids in $g_{DM}$ to extract the constraint. 

For reasons of speed, we used Pythia6 to decay the mediators \texttt{dr}, \texttt{sr}, which unfortunately forces the decay to be exactly on-shell.

Combining all the above samples requires some care when the width of the squark is large.  The reason is that vetoing internal on-shell squarks in the second and third sample technically removes the region of phase space corresponding to the squark momentum in the interval $\mathcal{I}\equiv m_M \pm \texttt{bwcutoff} \cdot \Gamma_M$, where \texttt{bwcutoff} is a Madgraph parameter whose default is 15. While this default value is adequate for a narrow width particle, where one is free to choose a number significantly larger than one while not affecting the phase space for samples two and three, in the case of a broad  particle, in order not to introduce significant kinematic distortions due to the on-shell decays, we chose \texttt{bwcutoff} such that $\texttt{bwcutoff} \cdot \Gamma_M$ is not much larger than ${\cal O}(50\,{\rm GeV})$, therefore often forcing a particular kinematic configuration not far from the peak of the resonance into sample 1 or 2.  This implies a contamination of on-shell mediators in samples 2 and 3 and the breaking of factorization $\sigma \cdot BR$ for the rates of processes 1 and 2.  To compensate, we re-scale samples 1 and 2 to avoid double counting by a factor 
\begin{equation}
\mathcal{N}_\mathcal{I}=\frac{\int_{\mathcal{I}}  BW(x) dx}{\Gamma_M}, \label{eq:bw}
\end{equation} 
where $BW(x)$ is the Breit-Wigner distribution of the width, and the integration in the numerator goes from $-\texttt{bwcutoff} \cdot \Gamma_M$ to $\texttt{bwcutoff} \cdot \Gamma_M$, whereas the integration in the denominator is over the entire Breit-Wigner.
Thus the number of events in a given bin reads
\be
n_i = (\mathcal{N})^2 \sigma^1\, \varepsilon_1^i + \mathcal{N} \sigma^2\, \varepsilon_2^i +  \sigma^3\, \varepsilon_3^i,
\ee
where the $\epsilon^i$ are the i-th bin efficiencies times luminosity for the various samples. 
This is clearly an approximation, since in the full matrix element, the Breit-Wigner function is integrated together with additional terms depending on the off-shellness of the squark, which have been neglected in Eq.\ref{eq:bw}. However we checked in explicit cases, using analytic formulae, that this introduces up to ${\rm O}(30\%)$ deviations in the rates, which is within our uncertainties once translated into a bound on the DM coupling. 

%
%
\newpage

\section{\textbf{Experimental analysis}}\label{appendix:Analysis}

For completeness, in Tables~\ref{dijetsmet} and Table.~\ref{monojet}, we report all the analyses used in the present work.

\begin{table}[h!]
\scriptsize
\caption{Dijets+met Analysis}
\begin{center}
\begin{tabular}{|c| m{11cm} |c|c|c|}
\hline
Name & \centering{Description} & TeV  & $\text{fb}^{-1}$ &cit.\\
\hline

 \href{https://atlas.web.cern.ch/Atlas/GROUPS/PHYSICS/CONFNOTES/ATLAS-CONF-2013-047}{ATLAS-CONF-2013-047} & {Search for squarks and gluinos in events with jets and MET: 0 leptons + 2-6 jets + Etmiss} & 8 &  20.3 &\cite{TheATLAScollaboration:2013fha}\\
 \hline
 \href{http://atlas.web.cern.ch/Atlas/GROUPS/PHYSICS/CONFNOTES/ATLAS-CONF-2012-109}{ATLAS-CONF-2012-109} &  {Search for squarks and gluinos using final states with jets and missing transverse momentum: 0 leptons + $\geq$2-6 jets + Etmiss} &  8 & 5.8&~\cite{ATLAS:2012ona}\\
 \hline
 \href{https://twiki.cern.ch/twiki/bin/view/CMSPublic/PhysicsResultsSUS13012}{CMS-PAS-SUS-13-012} &  {Search for New Physics in the Multijets and Missing Momentum Final State} & 8  & 19.5&~\cite{CMS-PAS-SUS-13-012}\\
 \hline
 \href{https://twiki.cern.ch/twiki/bin/view/CMSPublic/PhysicsResultsSUS12028}{CMS-PAS-SUS-12-028} &  {Search for supersymmetry in final states with missing transverse energy and 0, 1, 2, 3, or at least 4 b-quark jets using the variable AlphaT }&  8 &11.7&\cite{CMS:zxa} \\
 \hline
 \href{https://atlas.web.cern.ch/Atlas/GROUPS/PHYSICS/CONFNOTES/ATLAS-CONF-2012-033/}{ATLAS-CONF-2012-033} &  {Search for squarks and gluinos using final states with jets and missing transverse momentum} & 7 & 4.7&\cite{ATLAS-CONF-2012-033} \\
 \hline
 \href{https://twiki.cern.ch/twiki/bin/view/CMSPublic/PhysicsResultsSUS11022}{CMS-PAS-SUS-11-022} & {Search for supersymmetery in final states with missing transverse momentum and 0, 1, 2, or ³3 b jets} & 7 &4.98&\cite{CMS:2012kba} \\
 \hline
 \href{https://twiki.cern.ch/twiki/bin/view/CMSPublic/PhysicsResultsSUS12011}{CMS-PAS-SUS-12-011} &  {Search for new physics in the multijets + missing transverse energy final state}& 7 &4.98&~\cite{Chatrchyan:2012lia}\\
\hline
 \end{tabular}
\end{center}
\label{dijetsmet}
\end{table}%
\normalsize

\begin{table}[h!]
\scriptsize
\caption{Monojet Analysis}
\begin{center}
\begin{tabular}{|c| m{11cm} |c|c|c|}
\hline
Name & \centering{Description} & TeV  & $\text{fb}^{-1}$ &cit.\\
\hline

 \href{https://atlas.web.cern.ch/Atlas/GROUPS/PHYSICS/CONFNOTES/ATLAS-CONF-2013-068/}{ATLAS-CONF-2013-068} & { Search for direct stop pair production with stop decaying to charm and neutralino: 0 leptons + mono-jet/c-jets + Etmiss} & 8 & 20.3&\cite{TheATLAScollaboration:2013aia} \\
 \hline
 \href{https://atlas.web.cern.ch/Atlas/GROUPS/PHYSICS/CONFNOTES/ATLAS-CONF-2012-147}{ATLAS-CONF-2012-147} &  {Search for dark matter and large extra dimensions in events with a jet and missing transverse momentum} & 8 & 10.5 &\cite{ATLAS:2012zim}\\
 \hline
 \href{http://cds.cern.ch/record/1525585?ln=en}{CMS-PAS-SUS-12-048} & {Search for new physics in monojet events} & 8 & 19.5 &\cite{CMS-PAS-EXO-12-048}\\
 \hline
 \href{https://atlas.web.cern.ch/Atlas/GROUPS/PHYSICS/CONFNOTES/ATLAS-CONF-2012-84}{ATLAS-CONF-2012-084} &  {Search for dark matter candidates and large extra dimensions in events with a jet and missing transverse momentum} & 7 & 4.7 &~\cite{ATLAS:2012ky} \\
 \hline
 \href{https://twiki.cern.ch/twiki/bin/view/CMSPublic/PhysicsResultsEXO11059}{CMS-PAS-SUS-11-059} & {Search for New Physics with a Monojet and Missing Transverse Energy} & 7 &1.1&~\cite{Chatrchyan:2012me}\\

\hline
\end{tabular}
\end{center}
\label{monojet}
\end{table}%
\normalsize

\newpage

%
%

\section{ \textbf{Statistical analysis: $CL_s$}}\label{appendix:CLS}

\begin{figure*}[htb]
\begin{center}
\subfigure[]{\includegraphics[scale=0.35]{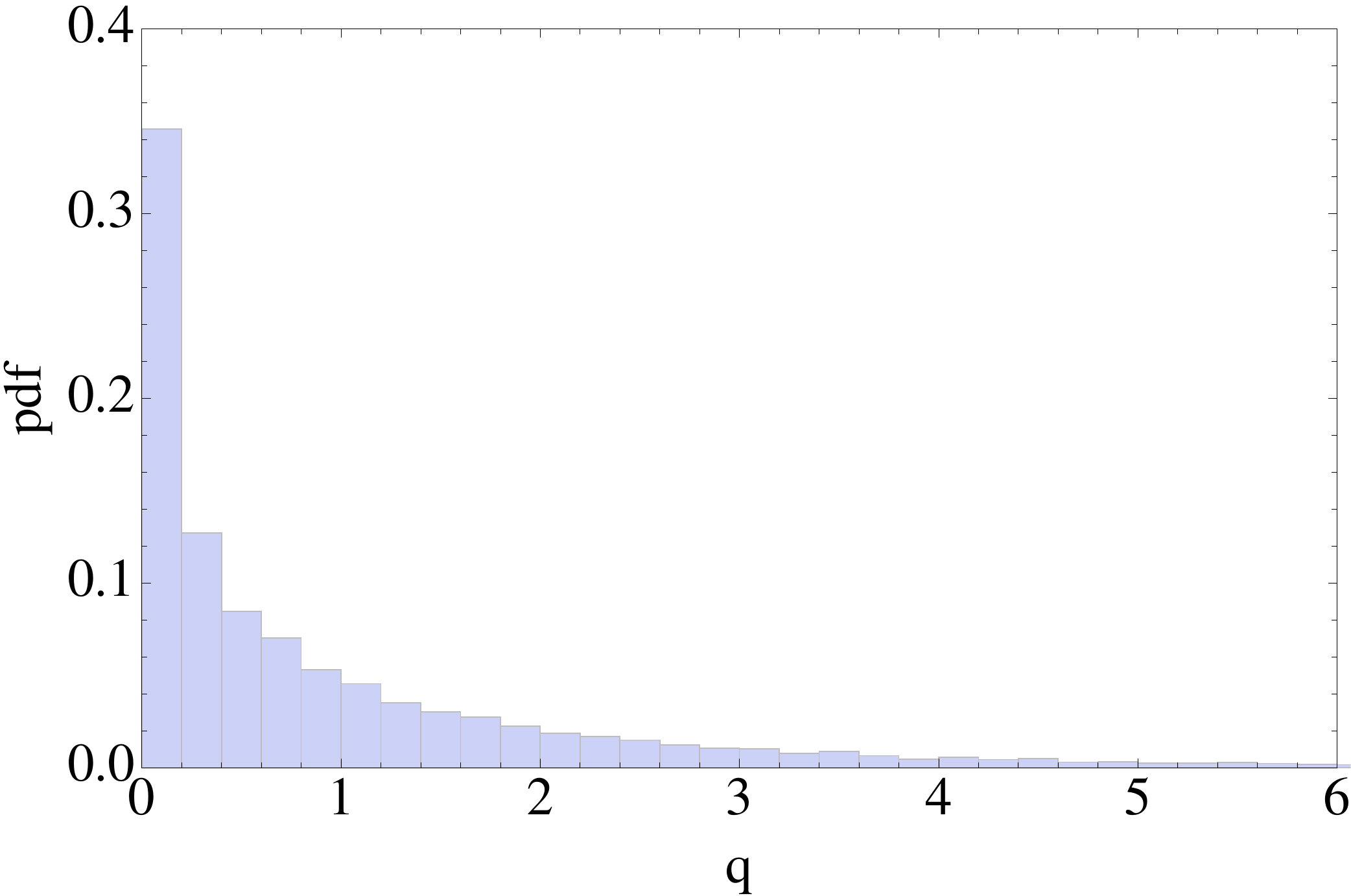}}
\subfigure[]{\includegraphics[scale=0.35]{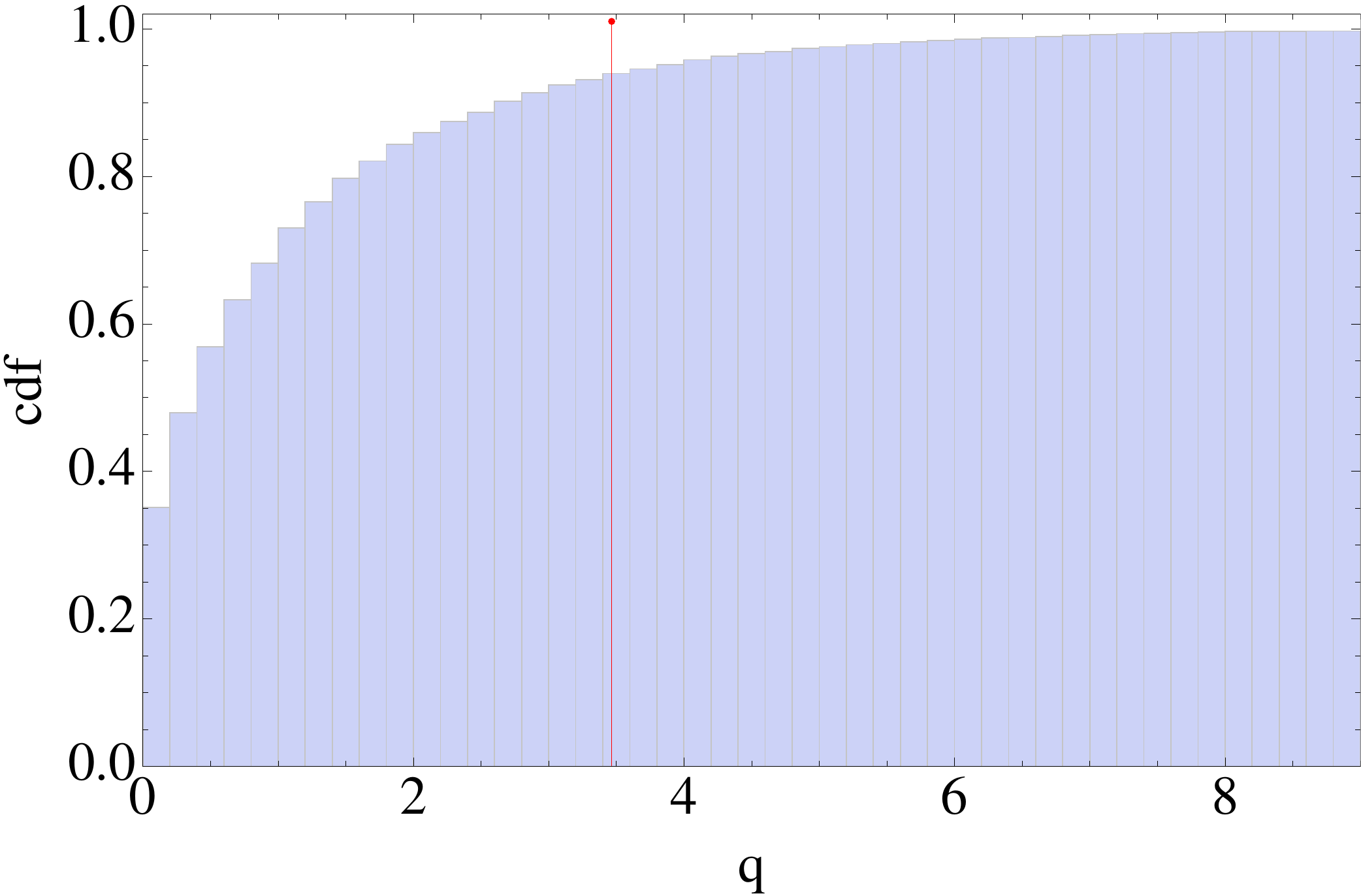}}
\caption{Probability distribution function (a) and cumulative distribution function (b) of the test statistic $q_{\text{b}+\text{s}}$ using data from~\cite{TheATLAScollaboration:2013aia}.}
\label{fig:test statistic pdf}
\end{center}
\end{figure*}

In this appendix we review the CLs method described in~\cite{Cowan:2010js, Mistlberger:2012rs}. In order to simplify the discussion we assume for the moment that we have only a single measurement: we will generalize to the multi-bin case later. We denote $n$ as the observed number of events in a given bin and we want to test the hypothesis that $n$ is made by a signal $n_\text{s}$ and a background $n_\text{b}$. The observed number of events will be distributed according to a Poisson statistics centered around the value $n_\text{s}+n_\text{b}$. As always, we do not know this central value, but can summarize the result of previous experience and simulations into a prior probability distribution for $n_\text{s}$ and $n_\text{b}$. For now, let us neglect the uncertainty connecting the signal and we focus on the background. In particular we will assume it follows a Gaussian distribution with central value $\mu_\text{b}$ and standard deviation $\sigma_\text{b}$. Hence the Likelihood for observing $n_{\text{obs}}$ is

\begin{eqnarray}\label{likelihood}
&& L_{\text{b}+\text{s}} (n_\text{b},n_\text{s})= \text{Poiss}(n, \mu_\text{s}+\mu_\text{b})  \mathcal{N}(\mu_\text{b},\sigma_\text{b}) \nonumber\\
&& \phantom{L_{\text{b}+\text{s}}} = \frac{e^{-\frac{\left(n_b-\mu _b\right){}^2}{2 \sigma _b^2}}}{\sqrt{2 \pi }
   \sigma _b} \frac{e^{-\mu _b-\mu _s} \left(\mu _b+\mu _s\right){}^n}{n!}.
\end{eqnarray}
In order to rule out the hypothesis of a given signal we define the test statistic
\begin{eqnarray}
&& q_{\text{b}+\text{s}} (n)= -2 \log\left(\frac{L_{\text{b}+\text{s}} (\hat{n}_\text{b},n_\text{s})}{L_{\text{b}+\text{s}} (\hat{\hat{n}}_\text{b},\hat{\hat{n}}_\text{s})}\right),
\end{eqnarray}
where $\hat{n}_\text{b}$ is the value that maximize s$L_{\text{b}+\text{s}} $ when $n_\text{s}$ is fixed, while $\hat{\hat{n}}_\text{b},\hat{\hat{n}}_\text{s}$ are the values that simultaneously maximize the Likelihood. Given the simple form of~\ref{likelihood} we can easily find the analytic solutions:
\begin{eqnarray}\label{maximization}
&& \hat{n}_\text{b} =\frac{1}{2} \left(\mu _b-\sigma _b^2-n_s+\sqrt{4 \sigma _b^2 n_{\text{obs}}+\left(\mu _b-\sigma_b^2+n_s\right){}^2}\right)\\
&& \hat{\hat{n}}_\text{b} = \mu_\text{b}\\
&& \hat{\hat{n}}_\text{s} = \max(n-\mu_\text{b},0). 
\end{eqnarray}
Note that the signal cannot be negative since it represents a number of events. Given a sample of values for $n$, distributed according to a mixture of Gaussian and Poisson statistics as in~\ref{likelihood},  one can compute the probability distribution function (pdf) of $q_{\text{b}+\text{s}}$. In Fig.~\ref{fig:test statistic pdf} we show the approximate pdf obtained using the M1 region data of the monojet ATLAS analysis~\cite{TheATLAScollaboration:2013aia}. 

Finally we define the quantity which measures the degree to which the signal plus background hypothesis is in agreement with the obesrvations:
\begin{eqnarray}
CL_{b+s}=P(q_{\text{b}+\text{s}} < q_{\text{b}+\text{s}}^{\text{obs}} ), \qquad \text{with} \qquad  q_{\text{b}+\text{s}}^{\text{obs}}= q_{\text{b}+\text{s}}(n=n_{\text{obs}}).
\end{eqnarray}

Exactly as above one can define a second test statistic, $q_{\text{b}} (n)$, this time making the assumption of no signal:
\begin{eqnarray}
&& q_{\text{b}} (n)= -2 \log\left(\frac{L_{\text{b}+\text{s}} (\hat{n}_\text{b},0)}{L_{\text{b}+\text{s}} (\hat{\hat{n}}_\text{b},\hat{\hat{n}}_\text{s})}\right),
\end{eqnarray}
where here the $n's$ are distributed with $n_s=0$. Hence we define
\begin{eqnarray}
CL_{b}=P(q_{\text{b}} < q_{\text{b}}^{\text{obs}} ), \qquad \text{with} \qquad  q_{\text{b}}^{\text{obs}}= q_{\text{b}}(n=n_{\text{obs}}).
\end{eqnarray}
We are finally ready to set our exclusion limit: the signal plus background hypothesis will be ruled out at a confidence level $\alpha$ if:
\begin{eqnarray}
CL_{s}  \equiv \frac{CL_{b+s}}{1-CL_b} = 1-\alpha .
\end{eqnarray}

The generalization to multiple bins is straightforward: the likelihood will be a product of the single bin likelihoods, and the test statistic $q$ will be $-2\log$ of this product. If the bins are completely independent, all the terms factorize in the maximization procedure and Eq.~\ref{maximization} still applies. On the other hand we are often interested in computing exclusion bounds on some fundamental parameter of the theory, whose variation will modify the signal in each bin in a well defined way. We can express this as $n_s=n_s(g)$, where $g$ is a coupling. The test statistic variable will then read
\begin{eqnarray}
&& q_{\text{b}+\text{s}} (n)= -2 \log\left(\frac{\prod_i L_{\text{b}+\text{s}} (\hat{n}^i_\text{b}, g)}{ \prod_iL_{\text{b}+\text{s}} (\hat{\hat{n}}^i_\text{b},\hat{\hat{g}})}\right).
\end{eqnarray}
When computing $\hat{\hat g}$, one should maximize the likelihood with respect to $g$ and all the $n^i_b$.  This might be hard to do analytically if the functions $n_s^i(g)$ are non trivial. Nevertheless, a necessary condition for the maximization is that $\partial L_{s+b}/\partial n_b^i =0$, which we can always solve due to the simple form of the likelihood. At this point one only needs to compute
\begin{eqnarray}
\max_g \,\,\prod_i L_{\text{b}+\text{s}} (\hat{n}^i_\text{b}(g), g),
\end{eqnarray}
which can in general be done numerically.

\newpage

\bibliography{monojet}
\bibliographystyle{apsrev4-1}

\end{document}